\documentclass[journal,twocolumn]{IEEEtran}
\usepackage{amsmath,graphicx}
\usepackage{color}
\usepackage{graphicx}
\usepackage{epstopdf}
\usepackage{amsmath}
\usepackage{amssymb}
\usepackage{mathrsfs}
\usepackage{hyperref}
\usepackage{tikz}
\usepackage{bm}
\usepackage[english]{babel}
\usepackage{cite}
\usepackage{rotfloat}
\usepackage{mathtools}
\usepackage[font=normalsize,labelfont=bf]{caption}
\usepackage{empheq}

\usepackage{amsmath}
\usepackage{makecell}
\usepackage{algorithm,algorithmic}
\usepackage{multirow}
\usepackage{subfigure}
\usepackage{booktabs}
\usepackage{colortbl}
\usepackage{multirow}
\usepackage{hhline}
\usepackage{stfloats}
\usepackage{multicol}
\usepackage{bbm}
\usepackage{cases}
\graphicspath{ {Figures/} }
\let\oldbibliography\thebibliography
\renewcommand{\thebibliography}[1]{%
  \oldbibliography{#1}%
  \setlength{\itemsep}{1.25pt}%
  \setlength{\baselineskip}{8.25pt}
  \setlength{\lineskiplimit}{-\maxdimen}
}

\newcommand{\T}{{\scriptscriptstyle\mathsf{T}}}
\renewcommand{\H}{{\scriptscriptstyle\mathsf{H}}}

\newsavebox{\foobox}

\definecolor{kugray5}{RGB}{224,224,224}

\usepackage[normalem]{ulem}
\newcommand\rsout{\bgroup\markoverwith
	{\textcolor{red}{\rule[0.5ex]{2pt}{0.8pt}}}\ULon}



\makeatletter
\newcommand{\ALOOP}[1]{\ALC@it\algorithmicloop\ #1%
	\begin{ALC@loop}}
	\newcommand{\ENDALOOP}{\end{ALC@loop}\ALC@it\algorithmicendloop}

\makeatother

\usepackage{etoolbox}
\let\mybibitem\bibitem
\renewcommand{\bibitem}[1]{%
	\ifstrequal{#1}{nature}
	{\color{blue}\mybibitem{#1}}
	{\color{black}\mybibitem{#1}}%
}

\graphicspath{ {Figures/} }

\newtheorem{theorem}{\textbf{Theorem}}
\newtheorem{remark}{\textbf{Remark}}

\newtheorem{proof}{Proof}

\newcommand{\epr}{\hfill\(\Box\)}

\DeclareCaptionLabelSeparator{periodspace}{.\quad}

\addto\captionsenglish{}
\newcommand\nbthis{\addtocounter{equation}{1}\tag{\theequation}}

\newcommand{\norm}[1]{\left\lVert#1\right\rVert} 
\newcommand{\fronorm}[1]{\left\lVert#1\right\rVert_{\mathcal{F}}} 
\newcommand{\abs}[1]{\left|#1\right|} 

\newcommand{\tr}[1]{\mathrm{tr}(#1)} 
\newcommand{\trt}[1]{\mathrm{tr}\left(#1\right)} 

\allowdisplaybreaks

\usepackage{setspace}

\newcommand{\mH}{{\mathbf{H}}} 

\newcommand{\mA}{{\mathbf{A}}}
\newcommand{\mW}{{\mathbf{W}}}

\newcommand{\mI}{\mathbf{I}}

\newcommand{\mX}{{\mathbf{X}}}

\newcommand{\mG}{{\mathbf{G}}}
\newcommand{\mF}{{\mathbf{F}}}
\newcommand{\mU}{{\mathbf{U}}}
\newcommand{\mV}{{\mathbf{V}}}

\newcommand{\mZ}{{\mathbf{Z}}}

\newcommand{\setC}{\mathbb{C}} 

\newcommand{\vs}{{\mathbf{s}}}

\newcommand{\vh}{{\mathbf{h}}}

\newcommand{\vw}{{\mathbf{w}}}
\newcommand{\va}{{\mathbf{a}}}

\def\b0{{\pmb{0}}}

\newcommand{\Nt}{N_\mathrm{t}}

\newcommand{\setK}{\mathcal{K}}

\newcommand{\Pt}{P_{\mathrm{t}}}
\newcommand{\noise}{\sigma_{\mathrm{n}}^2}

\begin{document}
	\setlength{\abovedisplayskip}{3.5pt}
	\setlength{\belowdisplayskip}{3.5pt}
	\title{Joint Communications and Sensing Hybrid Beamforming Design via Deep Unfolding}
	
	\author{Nhan Thanh Nguyen, \IEEEmembership{Member, IEEE},
		Ly V. Nguyen, \IEEEmembership{Member, IEEE},
		Nir Shlezinger, \IEEEmembership{Member, IEEE},\\
		Yonina C. Eldar, \IEEEmembership{Fellow, IEEE},
		A.~Lee~Swindlehurst, \IEEEmembership{Fellow, IEEE},
		and Markku Juntti, \IEEEmembership{Fellow, IEEE}
		\thanks{This research was supported by Academy of Finland under 6G Flagship (grant 346208), EERA Project (grant 332362), and Infotech Oulu.
			N. T. Nguyen and M. Juntti are with Centre for Wireless Communications, University of Oulu, P.O.Box 4500, FI-90014, Finland (e-mail: \{nhan.nguyen, markku.juntti\}@oulu.fi). L. V. Nguyen and A. L. Swindlehurst are with the Center for Pervasive Communications \& Computing, University of California, Irvine, CA, USA (email: vanln1, swindle@uci.edu). N. Shlezinger is with School of ECE, Ben-Gurion University of the Negev, Beer-Sheva, Israel (email: nirshl@bgu.ac.il). Y. C. Eldar is with Faculty of Math and CS, Weizmann Institute of Science, Rehovot, Israel (email: yonina.eldar@weizmann.ac.il).}\vspace{-0.25cm}
	}
	
	\maketitle
	
	\begin{abstract}
		Joint communications and sensing (JCAS) is envisioned as a key feature in future wireless communications networks. In massive MIMO-JCAS systems, hybrid beamforming (HBF) is typically employed to achieve satisfactory beamforming gains with reasonable hardware cost and power consumption. Due to the coupling of the analog and digital precoders in HBF and the dual objective in JCAS, JCAS-HBF design problems are very challenging and usually require highly complex algorithms. In this paper, we propose a fast HBF design for JCAS based on deep unfolding to optimize a tradeoff between the communications rate and sensing accuracy. We first derive closed-form expressions for the gradients {of the communications and sensing objectives} with respect to the precoders and demonstrate that {the magnitudes of} the gradients pertaining to the analog precoder are typically {smaller} than those associated with the digital precoder. Based on this observation, we propose a modified projected gradient ascent (PGA) method {with significantly improved convergence}. {We then develop a deep unfolded PGA scheme that efficiently optimizes the communications-sensing performance tradeoff with fast convergence thanks to the well-trained hyperparameters.} In doing so, we preserve the interpretability and flexibility of the optimizer while leveraging data to improve performance. Finally, our simulations demonstrate the potential of the proposed deep unfolded method, which achieves up to $33.5\%$ higher communications sum rate and $2.5$ dB lower beampattern error compared with the conventional design based on successive convex approximation and Riemannian manifold optimization. Furthermore, it attains up to a $65\%$ reduction in run time and computational complexity with respect to the PGA procedure without unfolding.
		
	\end{abstract}
	
	\begin{IEEEkeywords}
		Joint communications and sensing, dual-functional radar and communications, hybrid beamforming.
	\end{IEEEkeywords}
	\IEEEpeerreviewmaketitle
	
	\section{Introduction}
	
	Future wireless communications technologies such as evolving 6G systems will be required to meet increasingly demanding objectives. These include supporting massive numbers of static and mobile users, and enabling high-throughput low-latency communications in an energy-efficient manner. In addition to connectivity, 6G is expected to provide sensing and cognition capabilities~\cite{giordani2020toward}.    
	Various technological solutions are expected to be combined to satisfy these demands~\cite{samsung202065}. The millimeter-wave (mmWave) or Terahertz (THz) bands have been explored in this context~\cite{rappaport2019wireless, nguyen2022beam}. These bands provide large available bandwidth, thus, overcoming the spectral congestion of the conventional microwave and centimeter-wave (cmWave) communications bands. They can also inherently support high-resolution sensing~\cite{mishra2019toward}. 
	
	To generate directional beams and to cope with the harsh propagation profiles of high-frequency bands, wireless base stations (BSs) will employ large-scale massive multiple-input multiple-output (MIMO) arrays, typically implemented via hybrid beamforming (HBF) architectures to meet cost, power, and size constraints~\cite{molisch2017hybrid}. 
	Sensing capabilities can be enabled by high-frequency massive MIMO transceivers designed for dual communications and radar purposes~\cite{zhang2021overview}. This emerging concept of unifying communications and sensing is often called integrated sensing and communications (ISAC) \cite{ouyang2022performance, liu2022integrated} or joint communications and sensing (JCAS) \cite{zhang2018multibeam} which is the term used herein. Our focus in this paper is on the design of the transmitter for dual-functional radar-communications (DFRC) systems.
	
	Different forms of JCAS and DFRC systems have been proposed in the literature. Broadly speaking, the existing approaches can be classified based on their design focus~\cite{ma2020joint,zhang2021overview}. The first family of JCAS approaches are {\em radar-centric}, which build upon existing radar technologies while extending their operation to provide some communications capabilities, though typically with limited communications rates. This is often realized by integrating digital messages into the radar waveforms via index modulation~\cite{huang2020majorcom, ma2021spatial, ma2021frac} or by modulating the radar sidelobes~\cite{Hassanien2016Dual}. The alternative {\em communications-centric} approach aims at using conventional communications signals for probing the environment~\cite{Kumari2018IEEE80211ad}, though typically with limited sensing performance. The family of JCAS designs considered here employs {\em joint designs}, which enables balancing between the communications and sensing functionalities. 
	
	The spatial degrees-of-freedom provided by MIMO signaling can be exploited by JCAS systems based on joint designs, facilitating co-existence and resource sharing by beamforming~\cite{chepuri2022integrated}. However, the expected combination of JCAS systems operating at high frequencies using large-scale antenna arrays, particularly based on HBF, substantially complicates the beamforming design. Moreover, beamforming has to be established anew on each channel coherence interval, which at high frequencies can be on the order of less than a millisecond. This motivates the design of HBF JCAS systems that meet the requirements of both communications and sensing functionalities {with lower implementational complexity}, which is the focus of this paper.

	\subsection{Related Work}
	
	Transmit beamforming design for JCAS systems is the focus of growing attention in recent literature~\cite{liu2018mu,li2016optimum,liu2022transmit, pritzker2022transmit, liu2020joint, pritzker2021transmit,liu2018toward, tang2022mimo, wu2023constant, liu2021dual, liu2022joint,wu2022joint, johnston2022mimo, temiz2021dual, buzzi2019using, keskin2021mimo}. 
	Liu \textit{et al.} \cite{liu2018mu} proposed two JCAS strategies for multiuser MIMO systems with monostatic radar sensing where either two separated sub-arrays or a shared array at the BS are used. The shared array structure was demonstrated to give more reliable radar performance. Li \textit{et al.} in \cite{li2016optimum} designed the transmit covariance matrix so that the effective interference power at the radar receiver is minimized.  In \cite{johnston2022mimo}, Johnston \textit{et al.} developed radiated waveforms and receive filters, while Wu \textit{et al.} in \cite{wu2022joint} focused on optimizing data symbols to improve signal orthogonality in space and time. The studies in \cite{temiz2021dual, buzzi2019using, li2016optimum} showed that large-scale arrays can substantially mitigate the mutual interference between radar and communications.  Additional related works designed the overall transmit waveform as a superposition of the radar and communications waveforms \cite{liu2022transmit, pritzker2022transmit, liu2020joint, pritzker2021transmit} or using constant-modulus waveforms to achieve high energy efficiency at the nonlinear power amplifiers \cite{liu2018toward, tang2022mimo, wu2023constant, liu2021dual, liu2022joint}. These works consider fully digital MIMO architectures, which are expensive in terms of hardware and power consumption for high-frequency massive MIMO transceivers, while the resulting design is typically based on a complicated optimization procedure.
	
	
	HBF architectures realize large-scale MIMO transceivers in a cost-effective manner by delegating part of the signal processing to the analog domain. The most commonly considered implementation is based on analog phase shifter circuitry~\cite{molisch2017hybrid}, although alternative architectures employ vector modulators~\cite{zirtiloglu2022power}, metasurface antennas~\cite{shlezinger2021dynamic}, holographic surfaces~\cite{zhang2022holographic}, or variable amplifiers~\cite{gong2019rf}. Recent works have begun to explore HBF designs for JCAS~\cite{qi2022hybrid,wang2022partially,liyanaarachchi2021joint,barneto2021beamformer,cheng2022QoS,cheng2021hybrid_narrow,wang2022HBD_OFDM,islam2022integrated,cheng2021hybrid,Elbir2021HB_THz,zhang2018multibeam,Kaushik2021Hardware,Kaushik2022Green}. In particular, the work in~\cite{qi2022hybrid,wang2022partially,liyanaarachchi2021joint,barneto2021beamformer,cheng2022QoS} focuses on optimizing the radar performance under communications constraints. The approaches in~\cite{qi2022hybrid,cheng2022QoS} minimize the mean squared error between the transmit beampattern and a desired one subject to communications signal-to-interference-plus-noise ratio and data rate constraints,
	while alternative 
	metrics were used in \cite{wang2022partially,liyanaarachchi2021joint,barneto2021beamformer}. 
	The studies in~\cite{cheng2021hybrid_narrow} and~\cite{wang2022HBD_OFDM} follow a different design perspective that optimizes the communications performance under radar constraints. 
	In an effort to balance the radar and communications performance,  \cite{islam2022integrated} proposed to maximize the sum of the communications and radar signal-to-noise ratios (SNRs), while~\cite{cheng2021hybrid} optimized a weighted sum of the communications rate and radar beampattern matching error, and~\cite{Elbir2021HB_THz,liu2019hybrid} optimized the trade-off between the unconstrained communications beamformers and the desired radar beamformers. In~\cite{zhang2018multibeam}, Zhang \textit{et al.}  devised a multi-beam approach that employed a fixed sub-beam for communications and dynamic scanning sub-beams for the radar. Kaushik \textit{et al.} in \cite{Kaushik2021Hardware,Kaushik2022Green} considered the problem of RF chain selection to maximize energy efficiency. 
	While these works all consider HBF design for JCAS, they employ optimization procedures that are likely to be too lengthy to be implemented within a coherence interval, and that tend to involve many hyperparameters whose tuning, which has a crucial effect on the performance, is typically done manually.
	
	A data-driven approach to HBF design that avoids solving a complex optimization problem at runtime is to employ deep learning tools~\cite{shlezinger2023AI}.  
	The work of~\cite{Mateos2022EndtoEnd} and~\cite{muth2023Autoencoder} jointly learned the beamformers along with the target detection mapping and receiver processing in a deep end-to-end autoencoder model, while focusing on fully digital MIMO with a single receiver. 
	Xu \textit{et al.} \cite{xu2022deep} used deep reinforcement learning to design sparse transmit arrays with quantized phase shifters for HBF with a single RF chain, supporting a single user and while operating in either the radar or communications mode. Elbir {\it et al.}~\cite{elbir2021terahertz} trained two convolutional neural networks to estimate the direction to the radar targets, and considered a partially connected HBF in which the elements in each subarray are connected to the same phase shifter.

	\subsection{Motivations and Contributions}
	Employing deep learning for JSAC enables reliable  HBF designs to be carried out with low and fixed latency, unlike optimization-based approaches. However, existing designs employ deep neural networks (DNNs) based on black box architectures designed for conventional deep learning tasks, e.g., computer vision or pattern recognition. Consequently, unlike model-based optimizations, they are not interpretable and their training is often challenging and requires massive data sets. Such deep networks are trained for a specific setup, and deviations from that scenario, e.g., due to an increasing in the number of communications receivers, typically involves an architecture change and retraining. These limitations of optimization- and deep learning-based designs can be alleviated by {\em model-based deep learning} methodologies~\cite{shlezinger2020model}, and particularly {\em deep unfolding}~\cite{monga2021algorithm}. In deep unfolding, learning tools are employed to enhance the operation of an iterative optimizer that employs a fixed number of iterations~\cite{shlezinger2022model}. While deep unfolding methods have recently been shown to notably facilitate rapid HBF design for wireless communications~\cite{lavi2023learn,nguyen2023deep}, their application for JCAS has not yet been explored.
	
	In this paper, we propose an HBF design for JCAS systems based on deep unfolding. Our approach shares the interpretability and flexibility of optimization-based designs, along with low latency inference and the leveraging of data to improve the deep learning performance. In particular, we model the beamforming optimization problem using an objective that accommodates a  tradeoff between communications rate and matching a desired radar transmit beampattern as in~\cite{liu2018mu}, while bconstrained to the HBF architecture. We then formulate a candidate iterative solver for the optimization problem based on projected gradient ascent (PGA) with a dedicated initialization. Finally, we force the iterative solver to operate reliably within a fixed number of iterations by converting it into a trainable discriminative machine learning model~\cite{shlezinger2022discriminative}, whose trainable parameters are the hyperparameters of each PGA iteration. We also propose a training scheme that tunes these parameters using data in an unsupervised manner. Our main contributions are summarized as follows:
	\begin{itemize}
		\item We propose a novel HBF design for JCAS {\em transmission} based on deep unfolding of the PGA steps for optimizing a given tradeoff between the communications rate and the deviation from a desired transmit beampattern.
		\item By deriving the gradients used by PGA in closed form, we show that the {magnitudes of the gradients} of the objective with respect to the analog precoders are typically smaller than those corresponding to the digital precoder. Based on this observation, we alter PGA {to improve the convergence}.
		\item  We use deep learning tools to leverage data to tune the hyperparameters of the modified PGA algorithm to maximize the tradeoff objective within a given number of iterations. By doing so, we preserve the interpretability and flexibility of the optimizer while leveraging data to improve performance. 
		\item We extensively evaluate the proposed HBF design using various simulation studies. We demonstrate the gains of our proposed deep unfolded method in rapidly tuning hybrid precoders while simultaneously achieving significantly improved communications and sensing performance compared to the conventional iterative optimization schemes, including PGA without unfolding and the combined successive convex approximation (SCA) and Riemannian manifold optimization (ManOpt).
	\end{itemize}\vspace{-0.25cm}
	
	\subsection{Paper Organization and Notations}
	The rest of the paper is organized as follows. In Section \ref{sec_system_model}, we present the signal and channel models and the considered design problems. Section \ref{sec_proposed_scheme} details the proposed PGA and unfolded PGA schemes for JCAS-HBF. Numerical results are given and discussed in Section \ref{sec_simulation}. Finally, Section \ref{sec_conclusion} concludes the paper. 
	
	Throughout the paper, scalars, vectors, and matrices are denoted by lower-case, boldface lower-case, and boldface upper-case letters, respectively, while $[\mZ]_{i,j}$ is the $(i,j)$-th entry of matrix $\mZ$. We denote by $(\cdot)^\T$ and $(\cdot)^\H$ the transpose and the conjugate transpose operators, respectively, while $\abs{\cdot}$, $\norm{\cdot}$, and $\fronorm{\cdot}$ respectively denote the modulus of a complex number, the Euclidean norm of a vector, and the Frobenius norm of a matrix. We use $\mathcal{CN}(\mu, \sigma^2)$ to denote a complex normal distribution with mean $\mathbf{\mu}$ and variance $\sigma^2$, while $\mathcal{U}[a,b]$ denotes a uniform distribution over the range $[a,b]$.
	
	\section{Signal Model and Problem Formulation}
	\label{sec_system_model}
	

	\begin{figure}[t]
		\centering
		\includegraphics[scale=0.6]{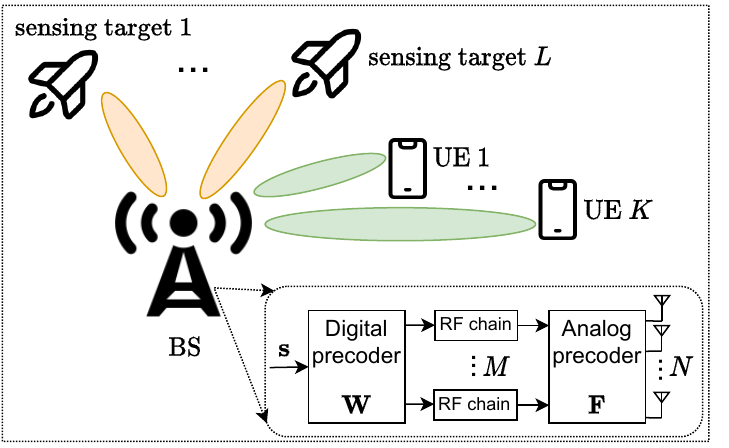}
		\caption{Illustration of the considered JCAS-HBF system.}
		\label{fig_HBF_system}
	\end{figure}
	
	\subsection{Signal Model}
	We consider a MIMO JCAS system in which a single BS equipped with $N$ antennas simultaneously transmits probing signals to $L$ sensing targets and data signals to $K$ single-antenna communications users (UEs), which then decode their intended data streams. The BS employs a  fully connected HBF architecture with phase shifter-based analog precoder $\mF \in \mathbb{C}^{N \times M}$ and digital precoder $\mW = [\vw_1, \vw_2, \ldots, \vw_K] \in \mathbb{C}^{M \times K}$, with power constraint $\fronorm{\mF \mW}^2 = \Pt$. Here, $M$ $(K \leq M \leq N)$ represents the number of RF chains at the BS. Let $\vs = [s_1, s_2, \ldots, s_K] \in {\mathbb{C}}^{K \times 1}$ be the transmitted symbol vector from the BS. Assuming that symbol $s_k$ and digital precoding vector $\vw_k$ are intended for UE $k$, the received signal at UE $k$ is given by 
	\begin{align*}
		&y_k = \underbrace{\vh_k^\H \mF \vw_k s_k}_{\text{desired signal}} + \underbrace{\vh_k^\H \sum_{\ell \neq k}^K \mF \vw_{\ell} s_\ell}_{\text{inter-user interference}} + \underbrace{n_k}_{\text{noise}}, \nbthis \label{processed_received_signal}
	\end{align*}
	where $n_k \sim \mathcal{CN}(0,\noise)$ is  additive white Gaussian noise, and $\vh_k \in \mathbb{C}^{N \times 1}$ is the channel vector from the BS to UE $k$. 
	
	We adopt the extended Saleh-Valenzuela  model \cite{yu2016alternating}:
	\begin{align}
		\vh_k = \sum_{q=1}^{Q} \alpha_{qk} \va(\phi_{qk}), \label{eq_channel_model}
	\end{align} 
	where $Q$ is the number of propagation paths, $\alpha_{qk}$ and $\phi_{qk}$ are the complex gain and angle of departure of the $q$-th path of the channel to UE $k$, respectively. In \eqref{eq_channel_model}, $\va(\phi_{qk}) \in \mathbb{C}^{N \times 1}$ denotes the transmit array response vectors, given as \cite{yu2016alternating, sohrabi2016hybrid}
	\begin{align}
		&\va(\phi_{qk}) =\frac{1}{\sqrt{N}}\Big[
		1,e^{j \pi \sin(\phi_{qk}) }, \dots,e^{j (N-1) \pi \sin(\phi_{qk})} \Big]^\T, \label{eq_array_response}
	\end{align}
	where we assume the deployment of a uniform linear array with half-wavelength antenna spacing. The assumption of a ULA is not strictly necessary, but it enables a simpler interpretation of the beampattern.
	
	\subsection{Problem Formulation} 
	
	Based on the signal model in \eqref{processed_received_signal}, the achievable sum rate over all the UEs is given as
	\begin{align*}
		&R = \sum_{k = 1}^K \log_2 \left( 1 + \frac{\abs{\vh_k^\H \mF \vw_k}^2 }{\sum_{\ell \neq k}^K \abs{\vh_k^\H \mF \vw_{\ell}}^2 + \noise} \right). \nbthis \label{eq_rate}
	\end{align*}
	The covariance matrix of the transmit signal vector is $\mF \mW \mW^\H \mF^\H$. Since the design of the beampattern is equivalent to the design of the covariance matrix of the transmit signals, the quality of the beampattern formed by the hybrid precoders $\{\mF, \mW\}$ can be measured by
	\begin{align*}
		\tau &\triangleq  \fronorm{ \mF \mW \mW^\H \mF^\H - \bm{\Psi} }^2 , \nbthis \label{eq_tau}
	\end{align*}
	where $\bm{\Psi} \in \setC^{N \times N}$ is the benchmark waveform matrix obtained by solving the following radar beampattern design problem \cite{liu2018mu, liu2018toward}
	\begin{subequations}
		\label{opt_beampattern}
		\begin{align*} 
			\quad \underset{\substack{ \alpha, \bm{\Psi} }}{\textrm{minimize}} \quad &  \sum_{t=1}^{T} \abs{\alpha \mathcal{P}_{\mathrm{d}}(\theta_t) - \bar{\va}(\theta_t)^\H  \bm{\Psi}  \bar{\va}(\theta_t)}^2 \nbthis \label{obj_func_beampattern} \\
			\textrm{subject to} \quad
			&[\bm{\Psi}]_{n,n} = \frac{\Pt}{N},\ \forall n \nbthis \label{cons_avr_power} \\
			&\bm{\Psi} \succeq \bm{0}, \bm{\Psi} = \bm{\Psi}^\H,  \nbthis \label{obj_cons_beampattern}
		\end{align*}
	\end{subequations}
	where $\{\theta_t\}_{t=1}^T$ defines a fine angular grid of $T$ angles that covers the detection range $[-\pi/2, \pi/2]$, $\bar{\va}(\theta_t) = [1,e^{j \pi \sin(\theta_t) }, \dots,e^{j (N-1) \pi \sin(\theta_t)}]$ is the steering vector of the transmit array, and $\alpha$ is a scaling factor \cite{liu2018mu}. Constraint \eqref{cons_avr_power} ensures that the waveform transmitted by different antennas has the same average transmit power \cite{liu2018mu}. This problem is convex and can be solved by standard tools such as CVX. Similar to prior work, we focus on the radar transmit beam constraints rather than the subsequent target detection and position estimation. The approach can be used for monostatic or multistatic radar setups.
	

	
	
	We are interested in a  JCAS-HBF design that maximizes the system sum rate constrained by the radar sensing metric $\tau$, the transmit power budget, as well as the hardware constraints of the analog beamformers:
	\begin{subequations}
		\label{opt_prob}
		\begin{align*} 
			\underset{\substack{ \mF, \mW }}{\textrm{maximize}} \quad & R \nbthis \label{obj_func} \\
			\textrm{subject to} \quad
			&\abs{[\mF]_{nm}} = 1, \forall n, m, \nbthis \label{cons_analog}\\
			& \fronorm{\mF \mW}^2 = \Pt, \nbthis \label{cons_power}\\
			&\tau \leq \tau_0, \nbthis \label{cons_beampattern}
		\end{align*}
	\end{subequations}
	where constraint \eqref{cons_analog} enforces the unit modulus of the analog precoding coefficients, \eqref{cons_power} is the power constraint, and \eqref{cons_beampattern} guarantees that the formed beampattern closely matches the benchmark $\va(\theta_t)^\H  \bm{\Psi}  \va(\theta_t)$. Problem \eqref{opt_prob} is nonconvex and therefore challenging to solve. Specifically, it inherits the constant-modulus constraints of HBF transceiver design \cite{nguyen2019unequally, yu2016alternating, sohrabi2016hybrid} and the strong coupling between the design variables $\mF$ and $\mW$ in the objective function \eqref{obj_func}, power constraint \eqref{cons_power}, and the radar constraint \eqref{cons_beampattern}.

	\section{Proposed Design}
	\label{sec_proposed_scheme}
	To address \eqref{opt_prob}, our main idea is to develop a multiobjective learning framework based on the PGA approach. This enables the simultaneous maximization of $R$ and minimization of $\tau$ via efficiently updating $\{\mF, \mW\}$. We first reformulate \eqref{opt_prob} as a multiobjective problem and develop the general PGA-based iterative solver below. Then, we propose an unfolded PGA algorithm to accelerate the convergence as well as to improve the performance of the design by leveraging data to cope with the non-convex nature of the problem.  
	
	\subsection{PGA Optimization Framework}
	
	We begin by reformulating \eqref{opt_prob} as
	\begin{subequations}
		\label{opt_prob_1}
		\begin{align*} 
			\underset{\substack{ \mF, \mW }}{\textrm{maximize}} \quad & R - \omega \tau \nbthis \label{obj_func_1} \\
			\textrm{subject to} \quad &\eqref{cons_analog}, \eqref{cons_power}. \nbthis \label{cons_PGA}
		\end{align*}
	\end{subequations}
	This reformulation integrates constraint \eqref{cons_beampattern} as a penalty term in the objective function \eqref{obj_func_1} with a regularization factor $\omega$. In principle, the coefficient $\omega$ needs to be dictated by the maximal beampattern deviation $\tau_0$. Here, we treat it as a given hyperparameter and study its effect in the sequel.
	
	In the case that the system employs the conventional fully digital beamformer, \eqref{opt_prob} can be solved via Riemannian manifold optimization, as in \cite{liu2018mu}.  However, as the analog and digital precoders are cast as design variables in \eqref{opt_prob} and \eqref{opt_prob_1}, the design in \cite{liu2018mu} is not readily applicable. 
	We propose leveraging the PGA method in combination with alternating optimization (AO) to solve \eqref{opt_prob_1}. Specifically, in each iteration, $\mF$ and $\mW$ are solved in an AO manner, i.e., one is solved while the other is kept fixed. The solutions to $\mF$ and $\mW$ are then projected onto the feasible space defined by \eqref{cons_analog} and \eqref{cons_power} via normalization.
	
	Specifically, for a fixed $\mW$, $\mF$ can be updated at the $(i+1)$-th iteration via projected gradient ascent steps, i.e., 
	\begin{align*}
		&\mF_{(i+1)} = \mF_{(i)} + \mu_{(i)} \left(\nabla_{\mF} R  - \omega \nabla_{\mF} \tau\right) \Big|_{\mF = \mF_{(i)}}, \nbthis \label{eq_F_PGA} \\
		&[\mF_{(i+1)}]_{nm} = \frac{[\mF_{(i+1)}]_{nm}}{\abs{[\mF_{(i+1)}]_{nm}}},\ \forall n, m, \nbthis \label{eq_projection_F}
	\end{align*}
	where $\nabla_{\mX} f$ is the gradient of a scalar-value function $f$ with respect to a complex matrix $\mX$. Similarly, given $\mF$,  $\mW$ can be updated at iteration $i+1$ as:
	\begin{align*}
		\mW_{(i+1)} &= \mW_{(i)} + \lambda_{(i)} \left(\nabla_{\mW} R  -  \omega \nabla_{\mW} \tau\right) \Big|_{\mW = \mW_{(i)}}, \nbthis \label{eq_W_PGA}\\
		\mW_{(i+1)} &= \frac{\Pt {\mW}_{(i+1)}}{\fronorm{{\mF}_{(i+1)} \mW_{(i+1)}}}. \nbthis \label{eq_projection_W}
	\end{align*}
	In this scheme, the closed-form gradients of $R$ and $\tau$ with respect to $\mF$ and $\mW$ are required. We derive these in the following theorems.

	\begin{theorem}
		\label{theo_gradient_rate}
		The gradients of $R$ with respect to $\mF$ and $\mW$ are given by \eqref{grad_rate_F} and \eqref{grad_rate_W}, respectively,
		\begin{figure*}[t]\vspace{-0.5cm}
			\begin{align*}
				\nabla_{\mF} R &= \sum_{k = 1}^K \frac{\tilde{\mH}_k \mF \mV}{\ln 2 \left(\tr{ \mF \mV \mF^\H \tilde{\mH}_k} + \noise \right)} - \sum_{k = 1}^K  \frac{\tilde{\mH}_k \mF \mV_{\bar{k}}}{\ln 2 \left(\tr{ \mF \mV_{\bar{k}} \mF^\H \tilde{\mH}_k} + \noise \right)}, \nbthis \label{grad_rate_F} \\
				\nabla_{\mW} R &= \sum_{k = 1}^K \frac{\bar{\mH}_k \mW}{ \ln 2 \left(\tr{\mW \mW^\H \bar{\mH}_k} + \noise \right)} - \sum_{k = 1}^K \frac{\bar{\mH}_k \mW_{\bar{k}}}{\ln 2  \left(\tr{\mW_{\bar{k}} \mW_{\bar{k}}^\H \bar{\mH}_k} + \noise \right)} , \nbthis \label{grad_rate_W}
			\end{align*}
			\hrule
			\vspace{-0.5cm}
		\end{figure*}
		where
		\begin{align*}
			&\mV \triangleq \mW \mW^\H \in \setC^{M \times M},\ \mV_{\bar{k}} \triangleq \mW_{\bar{k}} \mW_{\bar{k}}^H \in \setC^{M \times M}, \nbthis \label{def_V}\\
			&\tilde{\mH}_k \triangleq \vh_k \vh_k^\H \in \setC^{N \times N},\ \bar{\mH}_k \triangleq \mF^\H \tilde{\mH}_k \mF \in \setC^{M \times M}, \nbthis \label{def_Hmbar}
		\end{align*}
		and $\mW_{\bar{k}} \in \setC^{M \times K}$ is obtained by replacing the $k$-th column of $\mW$ with zeros.
	\end{theorem}
	\begin{proof}
		See Appendix \ref{app_proof_grad_rate}. \epr
	\end{proof}

	\begin{theorem}
		\label{theo_gradient_tau}
		The gradients of $\tau$ with respect to $\mF$ and $\mW$ are respectively given as
		\begin{align*}
			\nabla_{\mF} \tau &= 2 (\mF \mW \mW^\H \mF^\H - \bm{\Psi}) \mF \mW \mW^\H, \nbthis \label{grad_tau_F} \\
			\nabla_{\mW} \tau &= 2 \mF^\H (\mF \mW \mW^\H \mF^\H - \bm{\Psi}) \mF \mW. \nbthis \label{grad_tau_W}
		\end{align*}
	\end{theorem}
	
	\begin{proof}
		See Appendix \ref{app_proof_grad_tau}. \epr
	\end{proof}

	With the derived gradients, the update rules \eqref{eq_F_PGA} and \eqref{eq_W_PGA} are readily applied to obtain $\{\mF, \mW\}$. However, we found that such a straightforward application often yields poor convergence. This is because the gradients of $R$ and $\tau$ with respect to $\mF$ and $\mW$ are significantly {different in magnitude}, which affects their contributions to maximizing $R$ and minimizing $\tau$ at each iteration. Furthermore, recall that we are interested in fast solution for $\{\mF, \mW\}$, i.e., within a limited number of iterations. Consequently,  the step sizes $\{\mu_{(i)}, \lambda_{(i)}\}$ in \eqref{eq_F_PGA} and \eqref{eq_W_PGA} are critical factors affecting the performance achieved by the PGA method, and determining them is nontrivial. 
	While line search and backtracking \cite{boyd2004convex} can be employed to tune the step sizes at runtime, this would require excessive time and high computationalcomplexity since an additional optimization procedure must be tackled for each iteration. To improve convergence of the PGA procedure in \eqref{eq_F_PGA} and \eqref{eq_W_PGA} while enabling rapid tuning of the hybrid precoders, we first propose improved updating rules for $\mF$ and $\mW$ (Section~\ref{ssec:ImpPGA}), and then leverage data to tune the hyperparameters (step sizes) by incorporating them into a deep unfolded model (Section~\ref{ssec:UnfPGA}).
	
	\subsection{Proposed Improved PGA Procedure}
	\label{ssec:ImpPGA}
	
	We first analyze the unbalanced gradients of $R$ and $\tau$ with respect to $\mF$ and $\mW$ in the following remark.
	
	
	\begin{remark}
			\label{remark_grad_comparison}
			In \eqref{eq_F_PGA} and \eqref{eq_W_PGA}, the gradients $\nabla_{\mF} \tau$ and $\nabla_{\mW} \tau$ generally have significantly different magnitudes for large $N$:
			\begin{align*}
				\abs{\left[\nabla_{\mF} \tau\right]_{nm}} &\ll \abs{\left[\nabla_{\mW} \tau\right]_{mk}}, \nbthis \label{eq_grad_tau_compare}
			\end{align*}
			for $n = 1,\ldots,N$, $m = 1, \ldots, M$, and $k = 1,\ldots,K$.
		\end{remark}
		The comparison in \eqref{eq_grad_tau_compare} can be explained by the fact that the elements of $\mF$ are on the unit circle, and hence $\mF^\H \mF \approx N \mI_{M}$ for large $N$ since the diagonal elements are the result of a ``coherent'' sum of terms while the off-diagonal terms are not \cite{sohrabi2016hybrid, el2014spatially}. 
        Thus, from \eqref{grad_tau_F} and \eqref{grad_tau_W}, we have
		\begin{align*}
			\nabla_{\mF} \tau 
			&\approx 2 N \mF \mW \mW^\H \mW \mW^\H - 2 \bm{\Psi} \mF \mW \mW^\H, \nbthis \label{eq_grad_compare_2}\\
			\nabla_{\mW} \tau 
			&\approx 2 N^2 \mW \mW^\H \mW- 2 \mF^\H \bm{\Psi} \mF \mW \nbthis \label{eq_grad_compare_3}
		\end{align*}
		when $N$ is large. If $\Pt$ is fixed and the magnitudes of the entries of $\mW$ and $\bm{\Psi}$ are independent of $N$, $\abs{[\nabla_{\mF} \tau]_{nm}}$ and $\abs{[\nabla_{\mW} \tau]_{mk}}$ increase with a rate proportional to $N$ and $N^2$, respectively. However, it noted from  \eqref{cons_power} that
		\begin{align*}
			\Pt &= \fronorm{\mF \mW}^2 = \tr{\mF \mW \mW^\H \mF^\H}\\
			& = \tr{\mW \mW^\H \mF^\H \mF} \approx N \tr{\mW \mW^\H} = N \fronorm{\mW}^2,
		\end{align*}
		which yields $\fronorm{\mW}^2 \approx \Pt/N$. As a result, the magnitudes of the entries of $\mW$ generally decrease with a rate of $1/\sqrt{N}$. Furthermore, the diagonal elements of $\bm{\Psi}$ are equal to $\Pt/N$ while its non-diagonal elements are independent of $N$, as seen from \eqref{opt_beampattern}. Together, the above observations imply that $\abs{[\nabla_{\mF} \tau]_{nm}} \ll \abs{[\nabla_{\mW} \tau]_{mk}}$ 
        for large $N$. Furthermore, we also found in our numerical experiments that
		\begin{align*}
			\abs{\left[\nabla_{\mF} R \right]_{nm}} \ll \abs{\left[\nabla_{\mW} R \right]_{mk}},  \nbthis \label{eq_grad_R_compare}
		\end{align*}
		holds as well for large $N$, although the differences between $\abs{\left[\nabla_{\mF} R \right]_{nm}}$ and  $\abs{\left[\nabla_{\mW} R \right]_{mk}}$ are not as significant as those between $\abs{\left[\nabla_{\mF} \tau \right]_{nm}}$ and  $\abs{\left[\nabla_{\mW} \tau \right]_{mk}}$. 

  		\begin{figure}[t]
			\centering
			\includegraphics[scale=0.55]{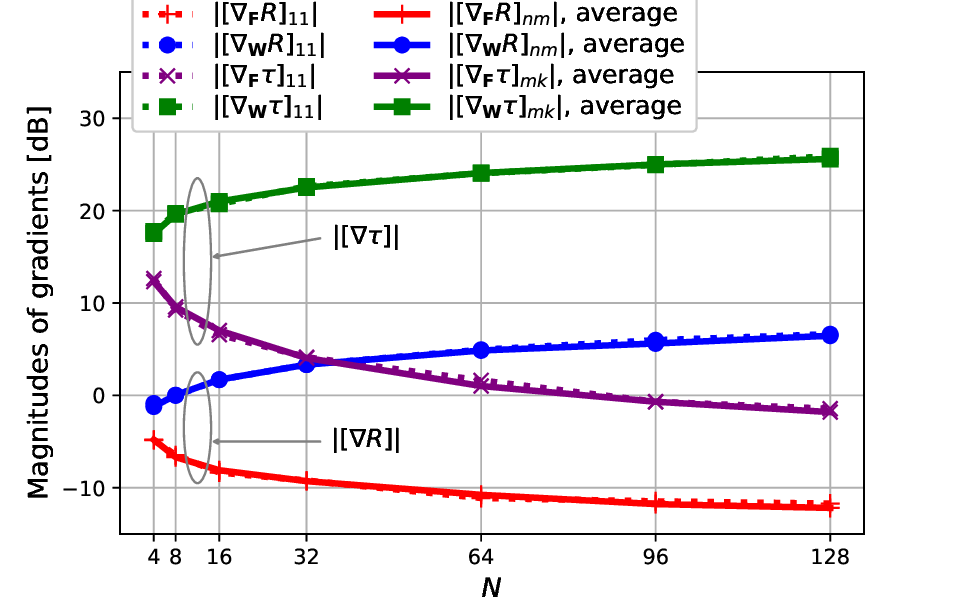}
			\caption{Comparison of the magnitudes of $\nabla_{\mF} R$, $\nabla_{\mW} R$, $\nabla_{\mF} \tau$, and $\nabla_{\mW} \tau$ with $N \in [4, 128]$, $K = M = 4$, and SNR $= 12$ dB.}
			\label{fig_grad_comparison}
		\end{figure}
		
		Fig.\ \ref{fig_grad_comparison} provides numerical evidence of the comparisons in \eqref{eq_grad_tau_compare} and \eqref{eq_grad_R_compare}, showing the magnitudes of the $(1,1)$-th element of $\{\nabla_{\mF} R, \nabla_{\mW} R, \nabla_{\mF} \tau, \nabla_{\mW} \tau\}$ as well the average magnitude of all the elements of these gradients for $100$ Monte Carlo simulations. The simulations assumed $N \in [4, 128]$, $K = M = 4$, SNR $= 12$ dB, and random but feasible $\{\mF, \mW\}$ so that the gradients are not affected by the optimality of the precoders. The results confirm that as $N$ increases, both $\abs{\left[\nabla_{\mW} R \right]_{mk}}$ and $\abs{\left[\nabla_{\mW} \tau \right]_{mk}}$ increase, while $\abs{\left[\nabla_{\mF} R \right]_{nm}}$ and $\abs{\left[\nabla_{\mF} \tau \right]_{nm}}$ decrease significantly. This holds true for both the magnitude of the $(1,1)$-th element as well as the average magnitude of all the elements of the considered gradients.

The observations in \eqref{eq_grad_tau_compare} and \eqref{eq_grad_R_compare} imply that in one iteration of the PGA procedure solving \eqref{opt_prob_1}, the update of $\mW$ is likely to be more dominant than that of  $\mF$, especially for the radar metric $\tau$. {If $\mF$ and $\mW$ were updated independently, the algorithm is expected to converge for a large enough number of iterations.} However, these variables are dependent and highly coupled in the objective function. Thus, even when $\mW$ is updated with a reasonable step size, its gradient still heavily depends on $\mF$, and vice versa. As a result, changes in one variable can directly affect the convergence behavior of the other. Therefore, the alternating updates between $\mF$ and $\mW$ means that a sub-optimal state for one of the variables negatively affects the other variable, and thus degrades the covergence of $R - \omega \tau$.

To overcome the above issue, we propose to modify the AO procedure, updating $\mF$ over multiple iterations before updating $\mW$ and imposing a weight $\eta$ on $\nabla_{\mW} \tau$. The approach enables $\mF$ to keep pace with $\mW$ during the PGA iterations. To describe the approach, let $I$ represent the number of outer iterations of the PGA scheme, and $J$ represent the number of inner iterations for updating $\mF$. For $i=0,\cdots,I$ and $j=0,\cdots,J-1$, $\mF$ is updated as:
\begin{subnumcases}{}
	\hat{\mF}_{(i,0)} = \mF_{(i)}, \label{eq_update_F} \\
	\hat{\mF}_{(i,j+1)} = \hat{\mF}_{(i,j)} + \mu_{(i,j)} \left(\nabla_{\mF} R  - \omega \nabla_{\mF} \tau\right) \Big|_{\mF = \hat{\mF}_{(i,j)}},  \label{eq_update_F_inner} \\
	\mF_{(i+1)} = \hat{\mF}_{(i,J)},  \label{eq_update_F_outer}
\end{subnumcases}
followed by the projection in \eqref{eq_projection_F}, where $\hat{\mF}_{(i,j)}$ and $\mu_{(i,j)}$ are respectively the precoder and step size in the $j$-th inner iterations of the $i$-th outer iteration, and $\mF_{(i)}$ is the final precoder obtained in the $i$-th outer iteration once all inner iterations have been completed. On the other hand, $\mW$ is updated as
\begin{align*}
	\mW_{(i+1)} &= \mW_{(i)} + \lambda_{(i)} \left( \nabla_{\mW} R  -  \omega \eta \nabla_{\mW} \tau\right) \Big|_{\mW = \mW_{(i)}}, \nbthis \label{eq_update_W}
\end{align*}
followed by the projection in \eqref{eq_projection_W}, where $\mW_{(i)}$ is the digital precoder obtained in the $i$-th outer iteration. 
Based on \eqref{eq_grad_compare_2}, \eqref{eq_grad_compare_3}, and via simulation, we found that $\eta = \frac{1}{N}$ leads to good convergence for PGA. According to our best knowledge, the modified updates for $\mF$ and $\mW$ in \eqref{eq_update_F}--\eqref{eq_update_W} have not been applied in existing AO and PGA-based HBF designs. Without the sensing objective, the conventional procedure in \eqref{eq_F_PGA}--\eqref{eq_projection_W} still leads to convergence of the communications rate \cite{agiv2022learn}. However, in the considered multi-objective JCAS-HBF problem where $\abs{\left[\nabla_{\mF} \tau\right]_{nm}} \ll \abs{\left[\nabla_{\mW} \tau\right]_{mk}}$, the modifications in \eqref{eq_update_F}--\eqref{eq_update_W} are required to significantly improve the convergence of $R - \omega \tau$.

\begin{figure}[t]
	\centering
	\includegraphics[scale=0.55]{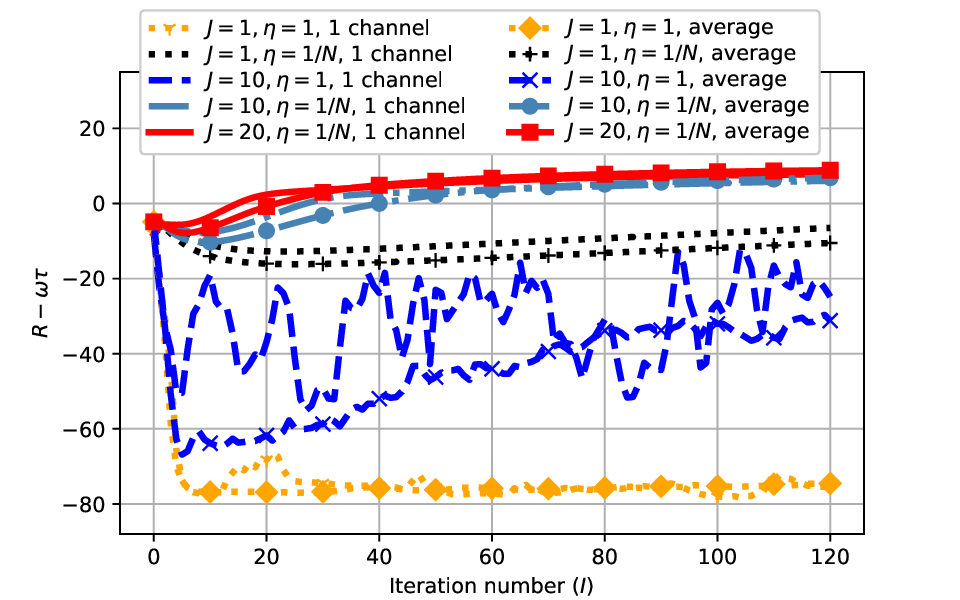}
	\caption{Convergence of the PGA algorithm with $N=32$, $K = M = 4$, $J = \{1, 10, 20\}$, SNR $= 12$ dB, and different weights for $\nabla_{\mW} \tau$.}
	\label{fig_comp_grad_weight}
\end{figure}

In Fig.\ \ref{fig_comp_grad_weight}, we depict the objective $R - \omega \tau$ over the PGA iterations for $J = \{1,10,20\}$, $\eta = \{1, \frac{1}{N}\}$, $N=32$, $K = M = 4$, and SNR $= 12$ dB. We fix the step sizes in all cases to $\mu_{(i,j)} = \mu_{(i)} = \lambda_{(i)} = 0.01, \forall i, j$, and we set $\omega = 0.3$ (we explain this choice in Section \ref{sec_simulation}). The convergence is shown for a single random channel realization as well as averaged over $100$ Monte Carlo simulations. {In each case we use the same initialization, which is specified in \eqref{eq_init} below.} It is seen that among the compared settings, $J = \{10,20\}$ and $\eta = \frac{1}{N}$ yield smooth convergence with increasing values for the objective. In contrast, setting $(J,\eta) = (1,1)$ results in non-increasing and unstable values for the objective over the iterations. This observation is consistent with the mutual effect between $\mF$ and $\mW$ discussed earlier. Setting either $J>1$ or $\eta=\frac{1}{N}$ improves convergence, but the value of the objective in these cases is still far worse than for $(J,\eta)=\left\{(10,\frac{1}{N}), (20,\frac{1}{N})\right\}$.

Although the proposed PGA mechanism can converge better than the conventional one, its convergence is still generally slow, especially for small $J$. In fact, the convergence speed largely depends on the step sizes. Next we propose an unfolded PGA framework with step sizes optimized via data-based training.

\subsection{Proposed Deep Unfolded PGA Model}
\label{ssec:UnfPGA}

The deep unfolding methodology encompasses several schemes that are based on converting an iterative optimizer with a fixed number of iterations into a trainable architecture that can be treated as a form of DNN~\cite{shlezinger2022model}. To preserve the interpretability and flexibility of PGA, we design our unfolded algorithm to fully preserve the PGA operation in Section \ref{ssec:ImpPGA}, while treating its hyperparameters, i.e., the step sizes $\{\mu_{(i,j)},\lambda_{(i)}\}_{i=0,j=0}^{I-1,J-1}$, as trainable parameters.

\subsubsection{Model Structure}

\begin{figure*}[t]
	\vspace{-0.5cm}
	\hspace{-0.3cm}
	\includegraphics[scale=0.60]{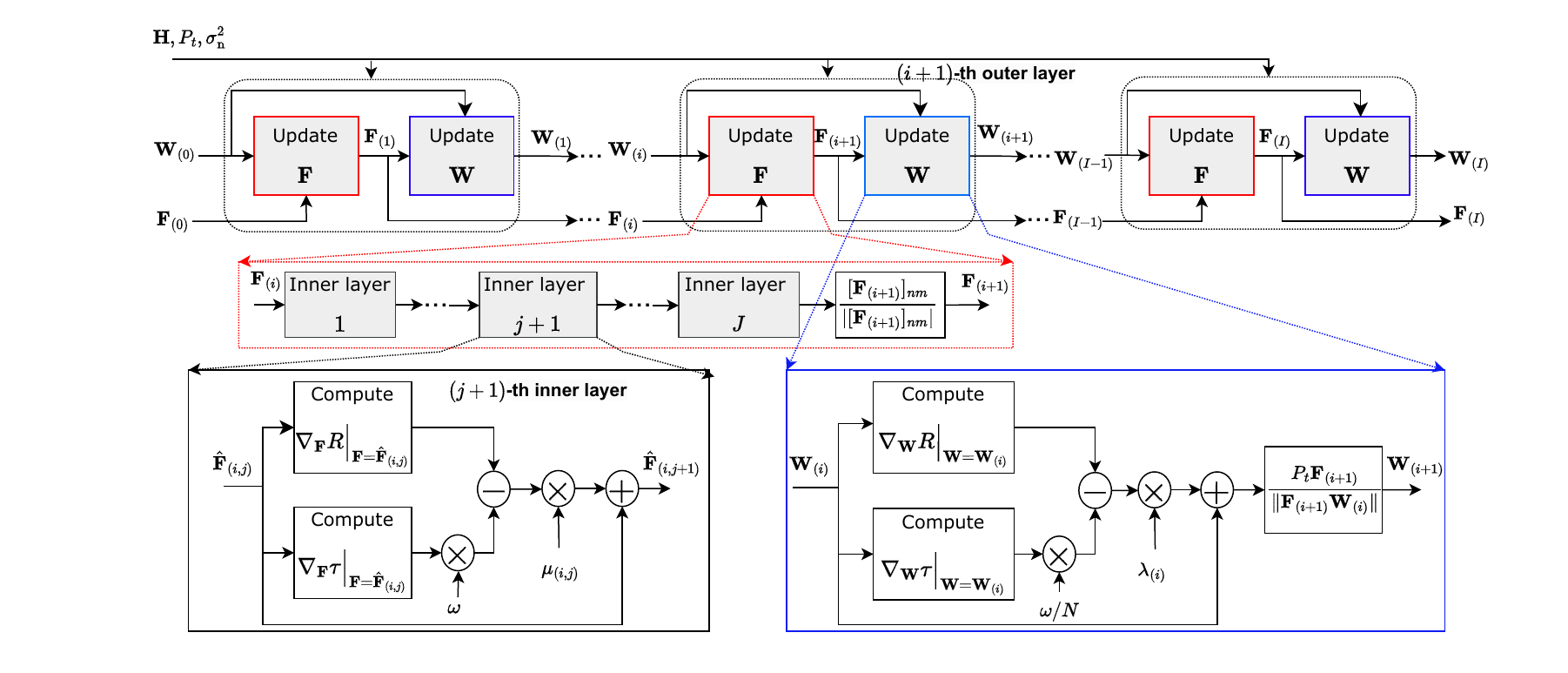}
	\caption{Illustration of the proposed unfolded PGA model.}\vspace{-0.5cm}
	\label{fig_unfolding_model}
\end{figure*}

Consider an unfolded PGA-based DNN of $I$ layers, unrolling the $I$ PGA iterations. The task of this model is to output feasible precoders $\{\mF, \mW\}$ with good communications and sensing performance, i.e., with $R - \omega\tau$ maximized. The unfolding mechanism maps an inner/outer iteration of the PGA procedure to an inner/outer layer of the unfolded PGA model. Therefore, we will still use subscripts $(i,j)$ to refer to the outer/inner layers when describing the unfolded PGA model. Furthermore, we denote $\bm{\mu} \triangleq \{\mu_{(i j)}\}_{i, j=0}^{I, J}$ and $\bm{\lambda} \triangleq \{\lambda_{(i)}\}_{i=0}^{I}$ for ease of exposition.

The unfolded PGA model, illustrated in Fig.\ \ref{fig_unfolding_model}, follows the updating process in \eqref{eq_update_F}--\eqref{eq_update_W}. It takes as input an initial guess $\{\mF_{(0)}, \mW_{(0)}\}$, the channel matrix $\mH = [\vh_1, \ldots, \vh_K]^\H$, the power budget $\Pt$, and the noise variance $\noise$, and it outputs $\{\mF_{(i)}, \mW_{(i)}\}$ over the outer layers $i = 1, \ldots, I$. Each outer layer includes a sub-network of $J$ layers to output $\mF_{(i)}$, mimicking the principle in \eqref{eq_update_F}--\eqref{eq_update_F_outer}. The operations inside each inner/outer layer include computing the gradients in \eqref{grad_rate_W}--\eqref{grad_tau_W} and applying the updating rules \eqref{eq_update_F}--\eqref{eq_update_W} and the projections \eqref{eq_projection_F} and \eqref{eq_projection_W}. The detailed operation of the unfolded PGA model will be further discussed in Section \ref{sec_overall_algorithm}.

\subsubsection{Training the Model}
As the unfolded architecture is derived from the optimization problem in \eqref{opt_prob_1}, it is trained to maximize $R - \omega \tau$. Accordingly, the loss function is set to
\begin{align*}
	&\mathcal{L}(\bm{\mu}, \bm{\lambda}) 
	= \omega \fronorm{ \mF_{(I)} \mW_{(I)} \mW_{(I)}^\H \mF_{(I)}^\H - \bm{\Psi} }^2\\
	&\qquad- \sum_{k = 1}^K \log_2 \left( 1 + \frac{\abs{\vh_k^\H \mF_{(I)} \vw_{k(I)}}^2 }{\sum_{\ell \neq k}^K \abs{\vh_k^\H \mF_{(I)} \vw_{j(I)}}^2 + \noise} \right) \nbthis \label{eq_loss_0}
\end{align*}
where \eqref{eq_loss_0} follows the original expressions of $R$ and $\tau$. The loss function $\mathcal{L}(\bm{\mu}, \bm{\lambda})$ enables training the model in an unsupervised manner. In particular, the data set is comprised of multiple channel realizations, and we boost the learned hyperparameters to be suitable for multiple SNRs by setting the noise power to $\noise = 1$ and randomly choosing $\Pt \in [\gamma_{\min}, \gamma_{\max}]$ dBW for each data sample so that the corresponding SNRs are in the range of interest $[\gamma_{\min}, \gamma_{\max}]$ dB. {We are interested in the moderate-to-high SNR regime, which is often required for both communications and radar sensing functions \cite{liu2018toward}. Therefore, we implement our simulations with a value for $\omega$ that ensures a good  tradeoff between $R$ and $\tau$ for moderate-to-high SNRs and treat it as a given hyperparameter during training of the unfolded model. Its chosen value and effects will be studied in Section \ref{sec_simulation}.} 

The loss $\mathcal{L}(\bm{\mu}, \bm{\lambda})$ is a function of the step sizes $\{\bm{\mu}, \bm{\lambda}\}$ because $\{\mF_{(I)}, \mW_{(I)}\}$ depends on $\{\mF_{(i)}\}_{i=0}^{I-1}$, $\{\mW_{(i)}\}_{i=0}^{I-1}$, and $\{\bm{\mu}, \bm{\lambda}\}$. The unfolded PGA model is trained to optimize $\{\bm{\mu}, \bm{\lambda}\}$ to achieve the best tradeoff within $I$ iterations. Furthermore, it is important to start the procedure with a good initial solution $\{\mF_{(0)}, \mW_{(0)}\}$. We discuss this issue in the next subsection.

\subsection{Overall Unfolded JCAS Algorithm}
\label{sec_overall_algorithm}
\subsubsection{Overall Algorithm}

\begin{algorithm}[t]
	\small
	\caption{Proposed Deep Unfolded PGA Algorithm}
	\label{alg_algorithm}
	\begin{algorithmic}[1]
		\REQUIRE $\mH$, $\Pt$, $\omega$, and the trained step sizes $\{\bm{\mu}, \bm{\lambda}\}$.
		\ENSURE $\mF$ and $\mW$
		\STATE \textbf{Initialization:} Generate $\{\mF_{(0)}, \mW_{(0)}\}$ based on \eqref{eq_init}.
		
		\FOR{$i = 0 \rightarrow I-1$}
		\STATE Set $\hat{\mF}_{(i,0)} = \mF_{(i)}$.
		\FOR{$j = 0 \rightarrow J - 1$}
		\STATE Obtain the gradients $\nabla_{\mF} R$ and $\nabla_{\mF} \tau$ at $(\mF, \mW) = (\hat{\mF}_{(i,j)}, \mW_{(i)})$ based on \eqref{grad_rate_F} and \eqref{grad_tau_F}.
		\STATE Obtain $\hat{\mF}_{(i,j+1)}$ based on \eqref{eq_update_F_inner}.
		\ENDFOR
		\STATE Set $\mF_{(i+1)} = \hat{\mF}_{(i,J)}$ and apply the projection in \eqref{eq_projection_F}.
		\STATE Obtain the gradients $\nabla_{\mW} R$ and $\nabla_{\mW} \tau$ at $(\mF, \mW) = \left(\mF_{(i+1)}, \mW_{(i)}\right)$ based on \eqref{grad_rate_W} and \eqref{grad_tau_W}.
		\STATE Obtain $\mW_{(i+1)}$ based on \eqref{eq_update_W} and apply the projection \eqref{eq_projection_W}.
		\ENDFOR
		\RETURN $\mF_{(I)}$ and $\mW_{(I)}$ as the solution to $\mF$ and $\mW$.
	\end{algorithmic}
\end{algorithm}

The proposed unfolded PGA algorithm for JCAS-HBF is outlined in Algorithm \ref{alg_algorithm}. The initial precoders $\{\mF_{(0)}, \mW_{(0)}\}$ are chosen as
\begin{align*}
	[\mF_{(0)}]_{nm} = e^{-j \vartheta_{nm}},\ 
	\mW_{(0)} = \mF_{(0)}^{\dagger} \mX_{\mathrm{ZF}}, \nbthis \label{eq_init}
\end{align*}
with $\mW_{(0)}$ normalized to satisfy \eqref{cons_power}, i.e., $\mW_{(0)} = \sqrt{\Pt} \mW_{(0)}/\fronorm{\mF_{(0)} \mW_{(0)}}$. In \eqref{eq_init}, $\vartheta_{nm}$ is the phase of the $(n,m)$-th entries of $\mG = [\vh_1, \ldots, \vh_K, \va(\theta_1), \ldots, \va(\theta_{M - K})]$, and $\mX_{\mathrm{ZF}} = \mG^\dagger$. Here, we assume that the number of RF chains is limited and smaller than the total number of UEs and targets, i.e., $M \leq K + L$. In mmWave massive MIMO systems, $M = K$ is a common setting for HBF architectures \cite{sohrabi2016hybrid}. In this case, we have $\mG = [\vh_1, \ldots, \vh_K]$, and $\mF_{(0)}$ becomes the same as the phased-ZF solution in \cite{liang2014low}. With \eqref{eq_init}, $\mF_{(0)}$ is aligned with the channels to harvest the large array gains. Furthermore, $\mW_{(0)}$ in \eqref{eq_init} is the constrained least-squares solution to the problem $\mathrm{min}_{\mW} \fronorm{\mF_{(0)} \mW - \mX_{\mathrm{ZF}}},~\mathrm{subject~to}~\eqref{cons_power}$. Therefore, the proposed input/initialization can provide good performance in multiuser massive MIMO systems, especially when $N$ is large. We will further verify this in Section \ref{sec_simulation}.

The unfolded model uses the trained step sizes $\{\bm{\mu}, \bm{\lambda}\}$ to perform the updates in \eqref{eq_update_F}--\eqref{eq_update_W} and the projections \eqref{eq_projection_F} and \eqref{eq_projection_W}, as outlined in steps 2--11 of Algorithm \ref{alg_algorithm}. Specifically, steps 3--8 compute the output $\mF_{(i+1)}$ over the $J$ layers. Then, $\mW_{(i+1)}$ is obtained in step 10 based on the updated $\mF_{(i+1)}$. The outcome of the algorithm is the final output of the unfolded PGA model.

\renewcommand{\arraystretch}{1.0}
\begin{table}[t!]
	\small
	\begin{center}
		\caption{Computational complexities involved in Algorithm \ref{alg_algorithm}.}
		\label{tab_complexity}
		\begin{tabular}{|c|c|}
			\hline
			Tasks & Complexities \\
			\hline
			\hline
			Compute $\nabla_{\mF} R$ & $\mathcal{O}(NM^2K)$ (per inner iteration/layer) \\
			\hline
			Compute $\nabla_{\mW} R$ & $\mathcal{O}(N^2K)$ (per inner iteration/layer) \\
			\hline 
			Compute $\nabla_{\mF} \tau$ & $\mathcal{O}(NMK)$  (per outer iteration/layer) \\
			\hline 
			Compute $\nabla_{\mF} \tau$ & $\mathcal{O}(N^2K)$ (per outer iteration/layer) \\
			\hline
			\hline
			Solve $\mF$ & $\mathcal{O}(IJN^2K)$ \\
			\hline 
			Solve $\mW$ & $\mathcal{O}(IN^2K)$ \\
			\hline 
			Overall algorithm & $\mathcal{O}(IJN^2K)$ \\
			\hline
		\end{tabular}
	\end{center}
\end{table}

\subsubsection{Complexity Analysis}
We end this section with a complexity analysis of the proposed JCAS-HBF design in Algorithm \ref{alg_algorithm}. First, we observe that $\mV$ and $\mV_{\bar{k}}$ are unchanged over $J$ inner iterations, while $\mW$ is of size $(M \times K)$ with $M, K \ll N$. Therefore, the main computational complexity of Algorithm \ref{alg_algorithm} comes from computing the gradients in \eqref{grad_rate_F}, \eqref{grad_tau_F}, \eqref{grad_rate_W}, and \eqref{grad_tau_W} in sequence, which are analyzed as follows.

The complexity of computing $\mathbf{\tilde{H}}_k\mathbf{F}$ in~\eqref{grad_rate_F} is only $\mathcal{O}(NM)$ because $\mathbf{\tilde{H}}_k\mathbf{F} = \vh_k \vh_k^\H\mathbf{F}$, which means that we can compute the term $\vh_k^\H\mathbf{F}$ first then perform a right-multiplication with $\vh_k$. The complexity in calculating $\mathbf{\tilde{H}}_k\mathbf{FV}$ is therefore $\mathcal{O}(NM^2)$, as a result of multiplying $\mathbf{\tilde{H}}_k\mathbf{F}$ with $\mV$. Computing $\operatorname{trace}\{\mF \mV \mF^\H \tilde{\mH}_k\}$ requires only $\mathcal{O}(NM)$ perations because $\mV \mF^\H \tilde{\mH}_k = (\mathbf{\tilde{H}}_k\mathbf{FV})^\H$, and $\mathbf{\tilde{H}}_k\mathbf{FV}$ has already been computed and the $\tr{\cdot}$ operator only requires the diagonal elements of its matrix argument. Thus, the complexity of the first summation term in \eqref{grad_rate_F} is $\mathcal{O}(NM^2K)$. Since the complexity of the two summation terms in \eqref{grad_rate_F} are the same, the total complexity in calculating \eqref{grad_rate_F} is still $\mathcal{O}(NM^2K)$. In \eqref{grad_tau_F}, the matrix $\mathbf{FW}$ is computed first and then is used to compute~\eqref{grad_tau_F}. Thus, the computational complexity of~\eqref{grad_tau_F} is $\mathcal{O}(N^2K)$, and combining the computational load of~\eqref{grad_rate_F} and~\eqref{grad_tau_F} results in an overall complexity of $\mathcal{O}(IJ\max(NM^2K,N^2K))$ to calculate the analog beaforming matrix $\mF$. 

Similarly, we can obtain the complexity of determining $\mathbf{W}$ as follows. Since $\bar{\mH}_k = \mF^\H \tilde{\mH}_k \mF = (\mathbf{h}_k^\H \mF)^\H(\mathbf{h}_k^\H \mF)$, we first compute $\mathbf{h}_k^\H \mF$ then use it to obtain $\bar{\mH}_k$ as $(\mathbf{h}_k^\H \mF)^\H(\mathbf{h}_k^\H \mF)$, with a complexity of $\mathcal{O}(NM)$. The complexity of calculating $\bar{\mH}_k\mW$ is thus $\mathcal{O}(\max(NM,M^2K))$ where $\mathcal{O}(M^2K)$ is the cost of multiplying $\bar{\mH}_k$ with $\mW$. With $\bar{\mH}_k\mW$ available, the complexity required to find $\tr{\mW \mW^\H \bar{\mH}_k}$ is only $\mathcal{O}(MK)$. The computational load required to compute the first term in \eqref{grad_rate_W} is thus $\mathcal{O}(\max(NMK,M^2K^2))$, which is also the total complexity required to calculate~\eqref{grad_rate_W} since the two terms of~\eqref{grad_rate_W} have the same complexity. Similar to~\eqref{grad_tau_F}, calculating~\eqref{grad_tau_W} requires $\mathcal{O}(N^2K)$ operations. Since $N \geq M$, we have $N^2K \geq NMK$, and therefore the total complexity in computing $\mathbf{W}$ is $\mathcal{O}(I\max(M^2K^2,N^2K))$.

Since $N \geq K$, we have $NM^2K \geq M^2K^2$, and so the complexity in solving for $\mathbf{F}$ dominates that for $\mathbf{W}$. Thus, the overal computational load for implementing the proposed deep unfolded PGA algorithm is $\mathcal{O}(IJ\max(NM^2K,N^2K))$ operations. Note that for HBF transceivers, it is generally true that $N \gg M, K$. Thus, we can approximate the component and overall complexities of Algorithm \ref{alg_algorithm} as in Table \ref{tab_complexity}. It is observed that for this algorithm, the per-iteration processing requires only a reasonable computational load of  $\mathcal{O}(N^2K)$ operations.

\begin{figure*}[!htb]
	\vspace{-0.8cm}
	\centering
	\subfigure[$N=32$]
	{
		\includegraphics[scale=0.55]{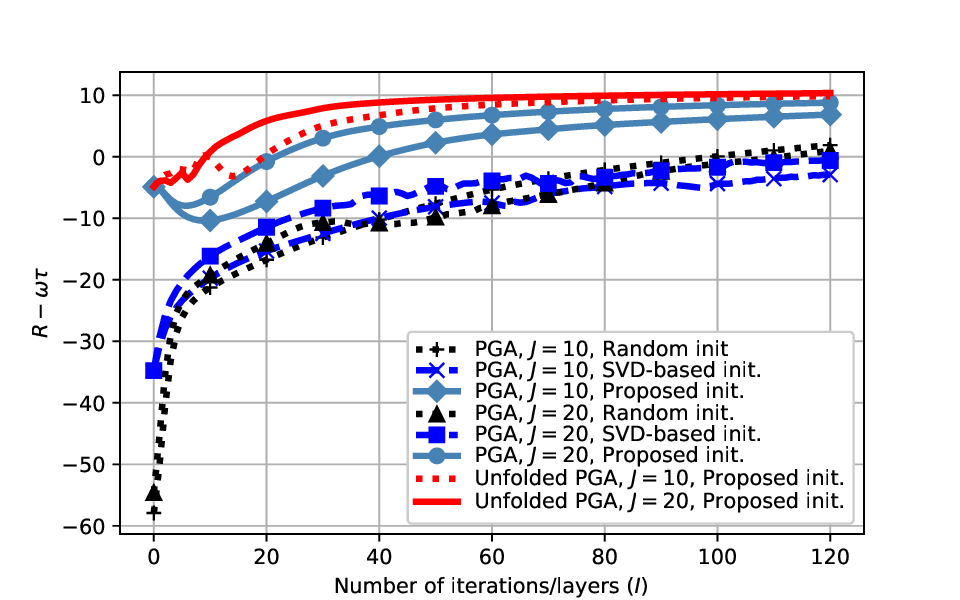}
		\label{fig_obj_iter_init_32}
	}
	\hspace{-0.5cm}
	\subfigure[$N=64$]
	{
		\includegraphics[scale=0.55]{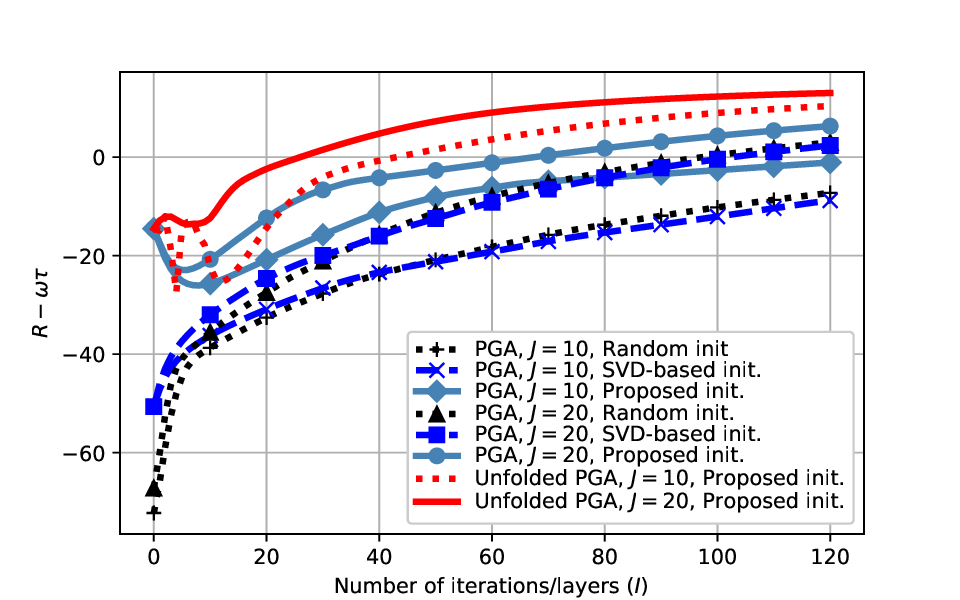}
		\label{fig_obj_iter_init_64}
	}
	\caption{Convergence of the PGA algorithm with $N=\{32, 64\}, K = M = 4, J = \{10, 20\}, \omega = 0.3$, SNR $= 12$ dB, and different initializations.}
	\label{fig_comp_conv_init}
\end{figure*}

\section{Simulation Results}
\label{sec_simulation}

Here we provide numerical results to demonstrate the performance of the proposed JCAS-HBF designs. We assume scenarios with $L=3$, $K = M = 4$ and $\Nt = \{32, 64\}$. To generate the channels in \eqref{eq_channel_model}, we set $Q = 10$, $\alpha_{qk} \sim \mathcal{CN}(0, 1)$, and $\theta_{qk} \sim \mathcal{U}(0, 2\pi)$ \cite{sohrabi2016hybrid}. The sensing targets are assumed to be located at angles $\theta_{\text{d}l} \in \{-60^{\circ}, 0^{\circ}, 60^{\circ}\}$, $l=1,\ldots,L$ and the corresponding desired beampattern is defined as \cite{cheng2021hybrid}
\begin{align*}
	\mathcal{P}_{\text{d}}(\theta_t) = \begin{cases}
		1,\ \theta_t \in [\theta_{\text{d}l} - \delta_{\theta}, \theta_{\text{d}l} + \delta_{\theta}]\\
		0,\ \text{otherwise}
	\end{cases}, \nbthis \label{eq_ideal_beampattern}
\end{align*}
where $\delta_{\theta} = 5$ is half the mainlobe beamwidth of $\mathcal{P}_{\text{d}}(\theta_t)$.

The deep unfolded PGA algorithm is implemented using Python with the Pytorch library. For the model training we set the decaying learning rate and initial learning rate to $0.97$ and $0.001$, respectively. The model is trained for $I=120$ and the SNR range $[\gamma_{\min}, \gamma_{\max}] = [0, 12]$ dB using the Adam optimizer with $1000$ channels over $100$ and $30$ epochs for $J = 1$ and $J = \{10,20\}$, respectively. We note that as long as $J$ is large enough, the proposed unfolded PGA model trained with $I \ll 120$ can still achieve satisfactory performance, as will be shown in Figs.\ \ref{fig_comp_conv_init}--\ref{fig_iter_64}. However, we provide the results for up to $I=120$ to show the long term behavior of the algorithms and to compare with the conventional PGA procedure over a large number of iterations. Unless otherwise stated, we set $\mu_{(0,0)} = \lambda_{(0)} = 0.01$, which are also used as the fixed step sizes for the PGA algorithm without unfolding. These are set based empirical observations. In the experiments whose results are reported in Figs.~\ref{fig_comp_grad_weight}--\ref{fig_SNR_64} we used the weighting coefficient $\omega = 0.3$, which was shown to offer a good communications--sensing performance tradeoff. We will further justify this setting by showing the results for various $\omega$ in Fig.\ \ref{fig_MSE_vs_omg_64}.



\subsection{Convergence and Complexity Discussion}

We have shown in Fig.\ \ref{fig_comp_grad_weight} that the conventional PGA approach with $(J,\eta)= (1,1)$ {and the initialization in \eqref{eq_init}} does not guarantee convergence. Therefore, we omit the results for this setting in the sequel. In Fig.\ \ref{fig_comp_conv_init}, we evaluate the effect of the initial solution/input $\{\mF_{(0)}, \mW_{(0)}\}$ in \eqref{eq_init} on the convergence of the (unfolded) PGA algorithm with $J = \{10,20\}$. For comparison, we consider the method in \cite{sohrabi2016hybrid}, which randomly generates $\mF_{(0)}$ and sets $\mW_{(0)} = \left( \mH \mF_{(0)}\right)^{\dagger}$ as the ZF solution based on the effective channel $\mH \mF_{(0)}$. For the second considered initialization approach, we assign the $M$ principal singular vectors of $\mH$ to $\mF_{(0)}$\cite{agiv2022learn} and set $\mW_{(0)}$ as in \eqref{eq_init}. For all these methods, $\{\mF_{(0)}, \mW_{(0)}\}$ are normalized to be feasible. We refer to these benchmarks in Fig.\ \ref{fig_comp_conv_init} as ``\textit{Random init}" and ``\textit{SVD-based init}," respectively.

It is observed from Fig.\ \ref{fig_comp_conv_init} that the proposed initialization substantially improves the convergence of PGA with and without unfolding. Specifically, $\{\mF_{(0)}, \mW_{(0)}\}$ in \eqref{eq_init} yields both a higher initial and final value for the PGA objective $R - \omega \tau$ than the other initializations. The SVD-based method yields a relatively good initial objective, but after a few iterations it behaves similarly to the random initialization, which has not converged after $120$ iterations. Furthermore, it is also seen that a larger $J$ leads to better convergence in all cases. For example, with $N=32$, setting $J=20$ allows the PGA approaches to obtain the peak of the objective about twice as fast as using $J=10$. Among the compared algorithms, the proposed unfolded PGA approach exhibits the best convergence in both Figs.\ \ref{fig_obj_iter_init_32} and \ref{fig_obj_iter_init_64}. With $N=64$, the unfolded PGA algorithm requires more iterations to converge, but its gain is more significant than with $N=32$.

The computational and runtime complexity reduction of the proposed PGA approaches are also observed in Fig.\ \ref{fig_comp_conv_init}. Here we compare (i) unfolded PGA with $J=20$, (ii) PGA with $J=20$, and (iii) PGA with $J=10$. All employ the proposed initialization. In Fig.\ \ref{fig_obj_iter_init_32}, these algorithms reach $R - \omega \tau = 0$ at $I \approx \{10, 20, 40\}$, respectively. This means that to achieve the peak value of the objective, approach (i) requires $n_{\mF} = IJ = 10 \times 20 = 200$ updates of $\mF$ and $n_{\mW} = I = 10$ updates of $\mW$. On the other hand, algorithms (ii) and (iii) require $(n_{\mF}, n_{\mW}) = (400, 20)$ and $(n_{\mF}, n_{\mW}) = (400, 40)$ updates, respectively. Similarly, we obtain $(n_{\mF}, n_{\mW})$ for $N=64$ in Fig.\ \ref{fig_obj_iter_init_64}, and the results are summarized in Table\ \ref{tab_no_iter}. We highlight $n_{\mF}$ since the time and computational complexity involved with finding $\mF$ dominates the overall algorithm. It is observed that with $J=20$, unfolded PGA achieves a reduction of approximately $\{50\%, 65\%\}$ in computational complexity and run time compared to PGA without unfolding in the scenarios $N = \{32, 64\}$, respectively. This is thanks to its optimized step sizes and the resulting small number of iterations required to achieve good performance.

\subsection{Communications and Sensing Performance}
\label{sec_R_tau}

\renewcommand{\arraystretch}{1.0}
\begin{table}[t!]
	\small
	\begin{center}
		\caption{Values of $(I,J,n_{\mF},n_{\mW})$ required by (unfolded) PGA to achieve $R - \omega \tau = 0$ in Fig.\ \ref{fig_comp_conv_init}. Here, $n_{\mF}$ and $n_{\mW}$ are the number of updates required for $\mF$ and $\mW$, respectively, with $n_{\mF} = IJ$, $n_{\mW}=I$.}
		\label{tab_no_iter}
		\begin{tabular}{|c|c|c|}
			\hline
			Schemes & $N=32$ & $N=64$\\
			\hline
			\hline
			\hspace{-0.2cm}Unfolded PGA $(J=20)$\hspace{-0.2cm} & $(10,20,\textbf{200},10)$ & $(25,20,\textbf{500},25)$\\
			\hline
			PGA $(J=20)$ & $(20,20,\textbf{400},20)$ & $(70,20,\textbf{1400},70)$\\
			\hline
			PGA $(J=10)$ & $(40,10,\textbf{400},40)$ & \hspace{-0.15cm}$(120,10,\textbf{1200},120)$\hspace{-0.2cm}\\
			\hline
		\end{tabular}
	\end{center}
\end{table}

\begin{figure*}[!htb]
	\vspace{-0.5cm}
	\hspace{-0.5cm}
	\subfigure[$R$ over iterations/layers]
	{
		\includegraphics[scale=0.41]{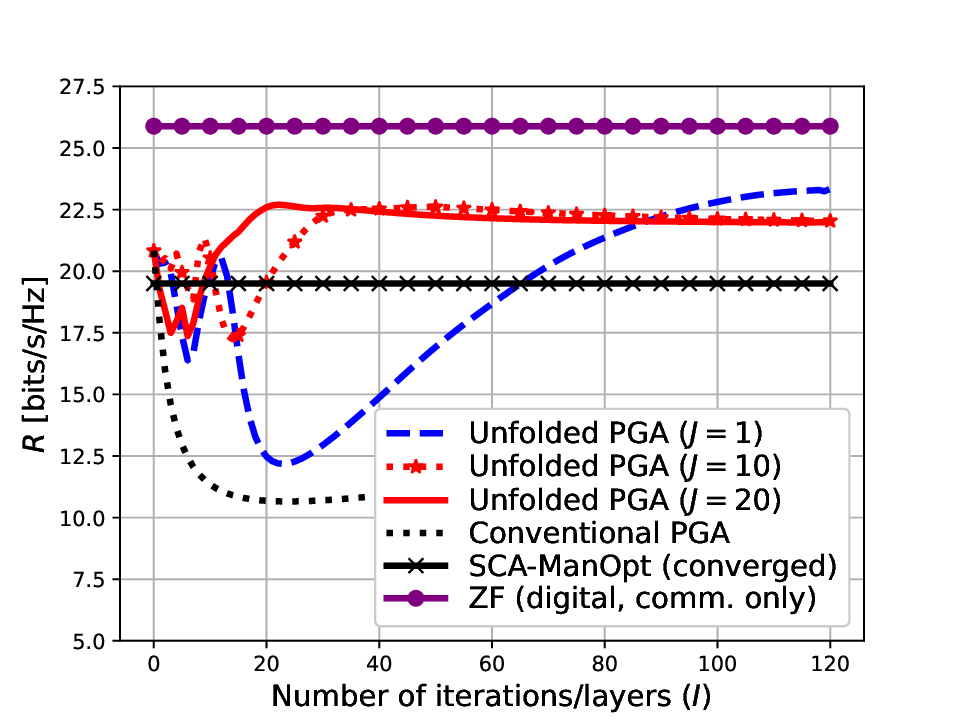}
		\label{fig_rate_vs_iter_32}
	}
	\hspace{-1cm}
	\subfigure[$\tau$ over iterations/layers]
	{
		\includegraphics[scale=0.41]{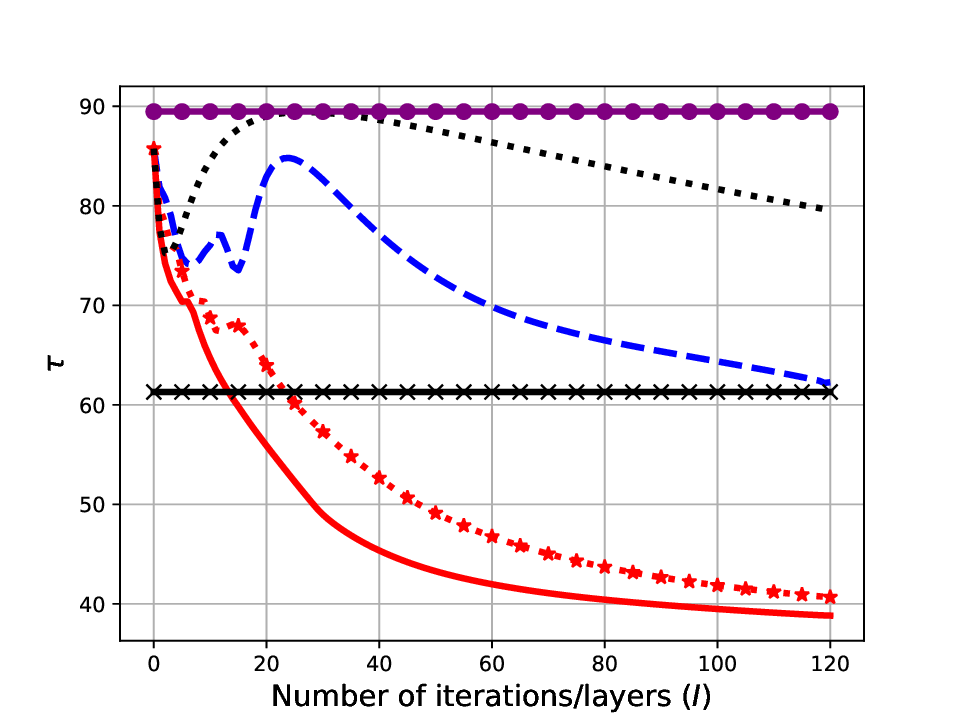}
		\label{fig_beampattern_error_vs_iter_32}
	}
	\hspace{-1cm}
	\subfigure[Sensing beampattern]
	{
		\includegraphics[scale=0.41]{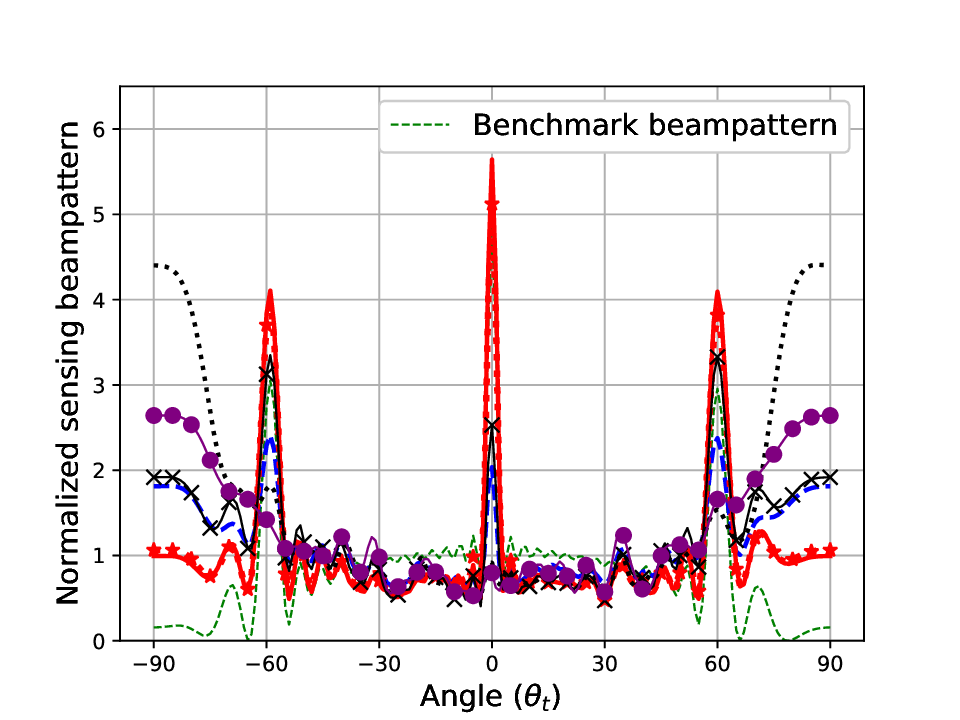}
		\label{fig_beampattern_32}
	}
	\caption{$R$, $\tau$, and beampattern of the considered approaches versus $I$, with $N=32$, $K = M = 4$, $J = \{1, 10, 20\}$, $\omega = 0.3$, and SNR $= 12$ dB.}
	\label{fig_iter_32}
\end{figure*}
\begin{figure*}[!htb]
	\vspace{-0.7cm}
	\hspace{-0.5cm}
	\subfigure[$R$ over iterations/layers]
	{
		\includegraphics[scale=0.41]{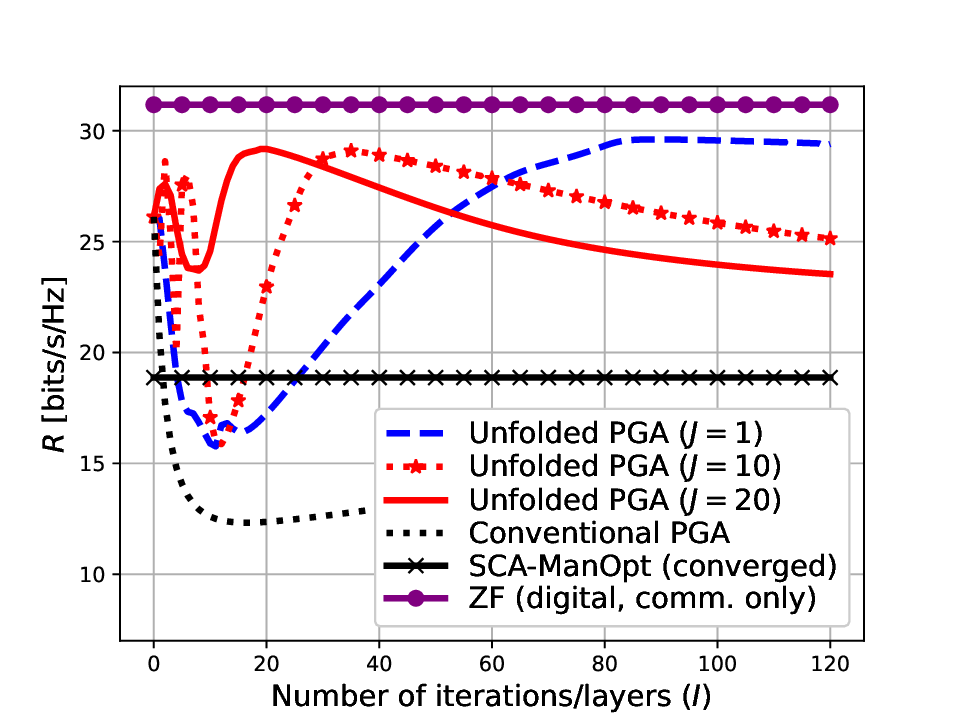}
		\label{fig_rate_vs_iter_64}
	}
	\hspace{-1cm}
	\subfigure[$\tau$ over iterations/layers]
	{
		\includegraphics[scale=0.41]{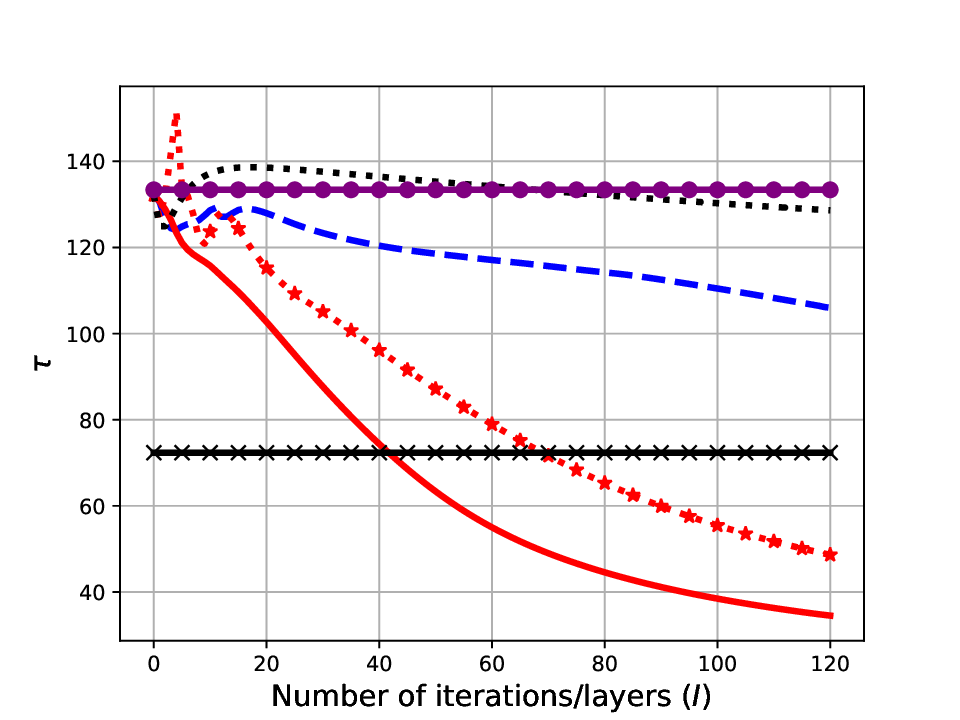}
		\label{fig_beampattern_error_vs_iter_64}
	}
	\hspace{-1cm}
	\subfigure[Sensing beampattern]
	{
		\includegraphics[scale=0.41]{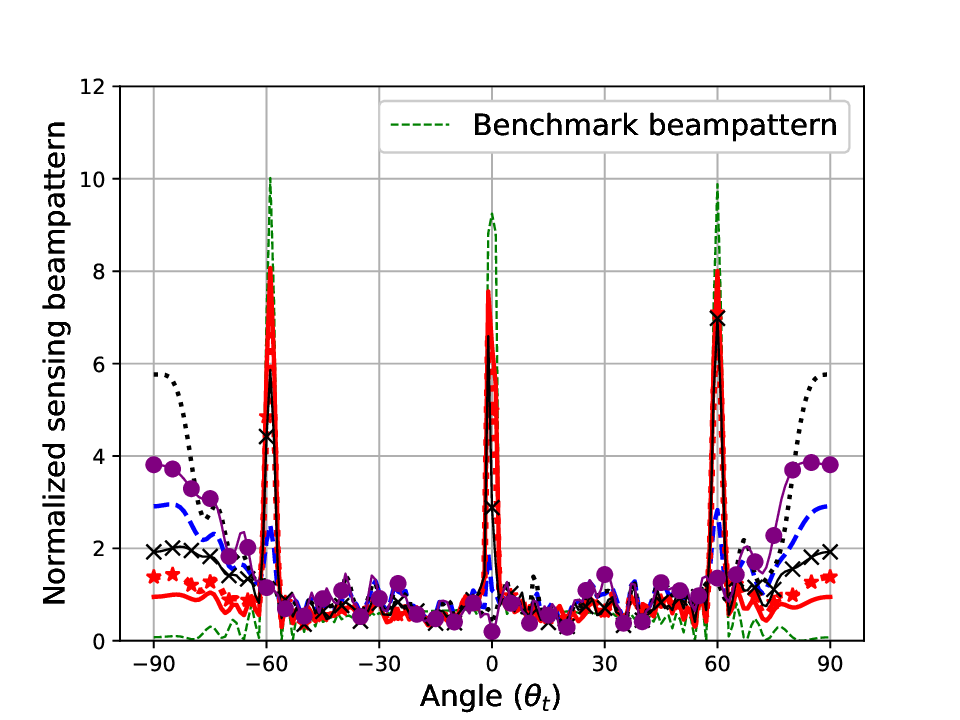}
		\label{fig_beampattern_64}
	}
	\caption{$R$, $\tau$, and beampattern of the considered approaches versus $I$, with $N=64$, $K = M = 4$, $J = \{1, 10, 20\}$, $\omega = 0.3$, and SNR $= 12$ dB.}\vspace{-0.25cm}
	\label{fig_iter_64}
\end{figure*}

We now focus on the communications and sensing performance of the proposed unfolded PGA algorithm.  For comparison, we consider the following approaches: (i)  \textit{conventional PGA} with fixed step sizes $\mu_{(0)} = \lambda_{(0)} = 0.01$, $J=1$, where we use $\eta = \frac{1}{N}$ instead of $\eta = 1$ to ensure smooth convergence. (ii) The JCAS-HBF design based on SCA and ManOpt (referred to as ``\textit{SCA-ManOpt}"). In this algorithm, an effective precoder $\mX^{\star}$ is first found that maximizes the communications sum rate via the iterative SCA approach \cite{tran2012fast}. Then, $\mX$ is obtained by maximizing $\rho \fronorm{ \mX - \mX^{\star} }^2 + (1-\rho) \fronorm{ \mX \mX^\H - \bm{\Psi} }^2$ with $\rho = 0.2$ \cite{liu2019hybrid, liu2018mu, liu2018toward} and $\{\mF, \mW\}$ are determined via matrix factorization \cite{yu2016alternating} leveraging the ManOpt scheme. We set the convergence tolerance to $\varepsilon = 10^{-3}$ for both the SCA and ManOpt procedures. (iii) The fully digital ZF beamformer in the communications-only system (referred to as ``\textit{ZF (digital, comm. only)}"). In downlink multiuser massive MIMO communications systems, the ZF beamformer performs near-optimally \cite{zhang2018performance}, and it provides an upper bound on the sum rate achieved by the JCAS-HBF approaches.

\begin{figure*}[!htb]
	\vspace{-0.5cm}
	\centering
	\subfigure[Sum rate]
	{
		\includegraphics[scale=0.55]{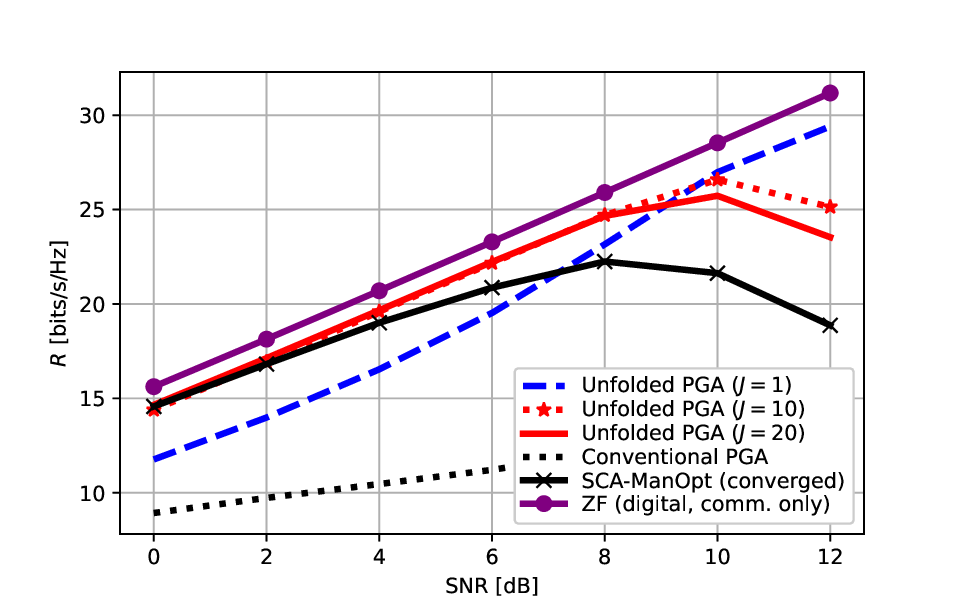}
		\label{fig_rate_vs_SNR_64}
	}
	\hspace{-0.5cm}
	\subfigure[Average beampattern MSE]
	{
		\includegraphics[scale=0.55]{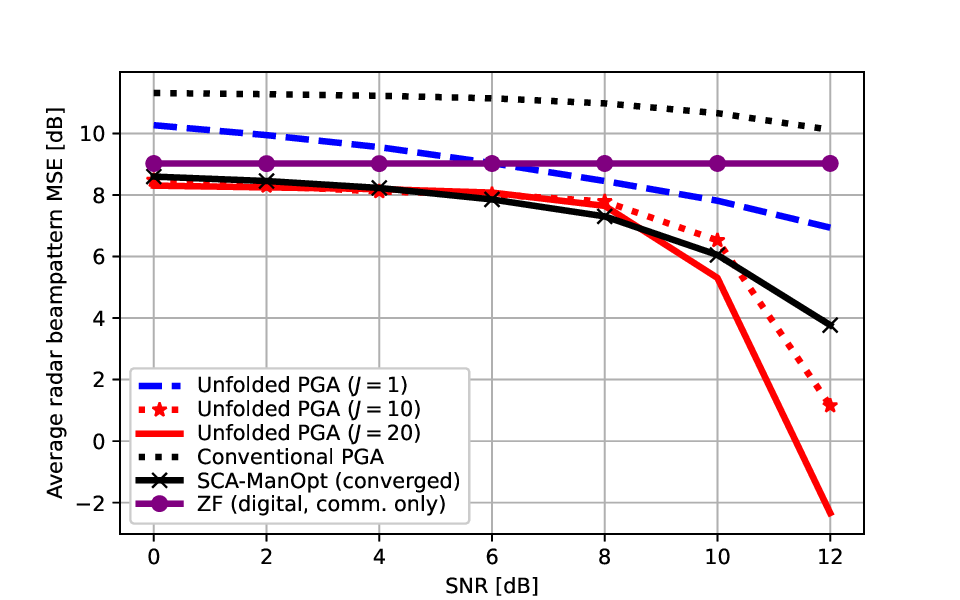}
		\label{fig_KSE_vs_SNR_64}
	}
	\caption{$R$ and the radar beampattern MSEs of the considered schemes versus SNRs with $N=64$, $K = M = 4$, $\omega = 0.3$, and $J = \{1, 10, 20\}$.}
	\label{fig_SNR_64}
\end{figure*}

In Figs.\ \ref{fig_iter_32} and \ref{fig_iter_64}, we present the communications and sensing metrics, i.e., $R$ and $\tau$, of the unfolded and conventional PGA algorithms versus the number of iterations/layers ($I$) and their resultant beampatterns. We also include the value of $R$ and $\tau$ of conventional ZF (digital, comm. only) and SCA-ManOpt at convergence for comparison. The same simulation parameters as in Fig.\ \ref{fig_comp_conv_init} are used. 
We observe the fluctuation of $R$ and $\tau$ for the first period of the PGA procedure (e.g., $I \in [0, 10]$, $J=20$ in Fig.\ \ref{fig_iter_64}). After the initial period $I \in [10, 20]$, $J=20$ in Fig.\ \ref{fig_iter_64}, $R$ increases while $\tau$ decreases rapidly. We note that the decrease in $R$, especially for large $I$, does not imply a performance loss. This variation is just the flexible adjustment of $\mF$ and $\mW$ in achieving a good communications-sensing performance tradeoff. Indeed, the objective $R - \omega \tau$ is still guaranteed to increase and converge, as seen in Fig.\ \ref{fig_comp_conv_init}. Comparing the unfolded PGA approaches, it is seen that different values for $J$ lead to different tradeoffs between $R$ and $\tau$, especially for $N=64$ in Fig.\ \ref{fig_iter_64}. As $I$ becomes sufficiently large, the case with $J=20$ yields a smaller $R$ but a much lower $\tau$ than what is obtained with $J = \{1, 10\}$, implying superior sensing performance. It is clear that the unfolded PGA algorithms with $J=\{10,20\}$ outperform their conventional PGA and the SCA-ManOpt counterparts in both communications and sensing performance. For example, in Figs.\ \ref{fig_rate_vs_iter_64} and \ref{fig_beampattern_error_vs_iter_64}, at $I=120$ the unfolded PGA algorithms with $J=\{10,20\}$ achieve a $\{33.2\%, 24.7\%\}$ improvement in $R$ and a $\{32.8\%, 52.3\%\}$ reduction in $\tau$, respectively, compared with SCA-ManOpt. They also perform close to the digital ZF precoder in terms of communications sum rate. 

We recall that $\tau$ measures the deviation of the designed beampattern from the benchmark, $\bm{\Psi}$, as shown in \eqref{eq_tau}. Because $\bm{\Psi}$ is optimized in \eqref{opt_beampattern} to achieve the desired sensing beampattern $\mathcal{P}_{\mathrm{d}}(\theta_t)$, the reduction in $\tau$ is equivalent to a better sensing beampattern. Indeed, it is observed from Figs.\ \ref{fig_beampattern_32} and \ref{fig_beampattern_64} that the sensing beampatterns obtained by the proposed unfolded PGA approaches fit the benchmark beampattern $\bar{\va}(\theta_t)^\H  \bm{\Psi}  \bar{\va}(\theta_t)$ in \eqref{obj_func_beampattern} the best. They have significantly higher peaks at the target angles $\{-60^{\circ}, 0^{\circ}, 60^{\circ}\}$ and lower side lobes compared to the beamptterns obtained with SCA-ManOpt and conventional PGA.

In Fig.\ \ref{fig_SNR_64}, we show the communications sum rate $R$ and the average radar beampattern mean square error (MSE) of the considered approches for $N=64$, $K = M = 4$, $\omega = 0.3$, $J = \{1, 10, 20\}$, and SNR $\in [0, 12]$ dB. The beampattern MSE is defined as $\text{MSE} = \frac{1}{T} \sum_{t=1}^{T} \abs{\mathcal{P}_{\text{d}}(\theta_t) - \bar{\va}^\H(\theta_t)  \bm{\Psi}  \bar{\va}(\theta_t)}^2$. We see from the figure that the proposed unfolded PGA algorithm with $J=\{5, 20\}$ performs close to the communications-only system with the fully digital ZF beamformer and outperforms SCA-ManOpt in terms of communications sum rates, while maintaining comparable or lower radar beampattern MSEs, especially at high SNR. For example, at SNR $= 12$ dB, the unfolded designs with $J=\{10, 20\}$ achieve about $\{33.2\%, 24.7\%\}$ higher sum rates and $\{2.5, 6\}$ dB lower MSEs compared with SCA-ManOpt, respectively. While the unfolded PGA employing $J=1$ can offer good communications performance at high SNR, its sensing performance is poor. Among the considered cases, conventional PGA with $J=1$ has the worst performance for both the communications and sensing operations. 

\subsection{Effects of $\omega$ on the JCAS-HBF Performance}

\begin{figure}[t]
	\hspace{-0.52cm}
	\includegraphics[scale=0.51]{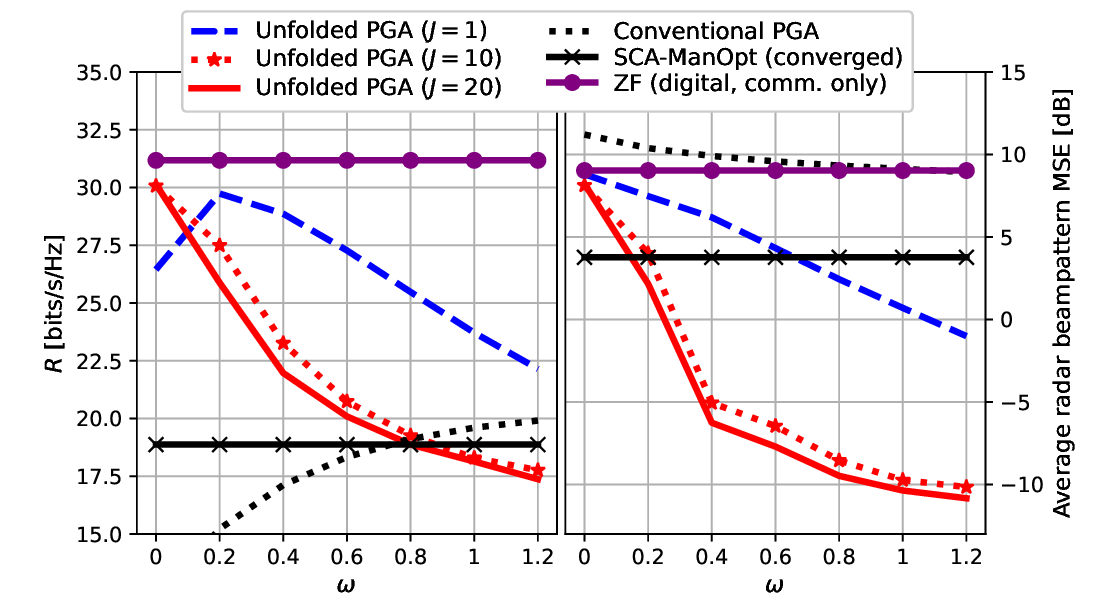}
	\caption{$R$ and the radar beampattern MSEs of the considered versus $\omega$ with $N=64, K = M = 4, J = \{1, 10, 20\}$, and SNR $= 12$ dB.}
	\label{fig_MSE_vs_omg_64}
\end{figure}

Finally, we investigate the effects of $\omega$ on the communications and sensing performance in Fig.\ \ref{fig_MSE_vs_omg_64}. For the proposed unfolded PGA approaches, we see that as $\omega$ increases, both $R$ and $\tau$ significantly decrease. To explain this, we revisit the objective function $R - \omega \tau$ in \eqref{opt_prob_1}, and we note that $R - \omega \tau \rightarrow - \omega \tau$ as $\omega \rightarrow \infty$. As a result, the PGA method tends to minimize $\tau$ rather than maximizing $R$ when $\omega$ is sufficiently large, and vice versa for $\omega \rightarrow 0$. To ensure a good communications-sensing performance tradeoff, $\omega$ should be chosen to balance the objectives $R$ and $\tau$, and this operating point can be tuned via simulation. It is seen in Fig.\ \ref{fig_MSE_vs_omg_64} that when $\omega = 0.3$, the unfolded PGA approaches with $J = \{10, 20\}$ can achieve better beampattern MSEs and a much higher sum rate than SCA-ManOpt. Furthermore, we also see from Figs.\ \ref{fig_iter_32} and \ref{fig_iter_64} that the relationship $R \approx 0.3 \tau$ holds for SCA-ManOpt at convergence. Therefore, we have set $\omega = 0.3$ in the previous simulations. However, as seen in Fig.\ \ref{fig_MSE_vs_omg_64}, different values of $\omega$ can be used depending on the JCAS design objectives. For example, in another radar-centric design aiming at high sensing accuracy, a large $\omega$ should be chosen. In contrast, in the communications-centric design considered in \eqref{opt_prob}, a moderate $\omega$ offers better communications performance.

\section{Conclusions}
\label{sec_conclusion}

We have studied multiuser massive MIMO JCAS systems with HBF transceiver architectures, aiming at maximizing the communications sum rate constraining the radar sensing beampattern accuracy. We reformulated the constrained problem into a multiobjective optimization that accounts for the tradeoff between the communications and sensing metrics. By analyzing the gradients of those metrics, we proposed effective updating rules for the analog and digital precoders to obtain smooth convergence of the PGA optimization. We further proposed an efficient unfolded PGA approach based on the deep unfolding technique, where the step sizes of the PGA approach are learned in an unsupervised manner. While both the proposed PGA algorithm without unfolding has reasonable computational complexity, the unfolded version is much faster with significantly reduced computational complexity thanks to its well-trained step sizes. Our extensive numerical results demonstrate that the unfolded PGA approach achieves significant improvements in communications and sensing performance with respect to conventional JCAS-HBF designs. Our future work will consider more practical partially connected HBF architectures and the wideband signal case where the analog array becomes frequency selective with beam squint.

\appendices

\section{Proof of Theorem \ref{theo_gradient_rate}}
\label{app_proof_grad_rate}

First, we rewrite the sum rate expression in \eqref{eq_rate} as
\begin{align*}
	R
	&= \sum_{k = 1}^K  \log_2 \left(\frac{\sum_{k = 1}^K \abs{\vh_k^\H \mF \vw_k}^2  + \noise }{\sum_{\ell \in \setK \backslash k}  \abs{\vh_k^\H \mF \vw_{\ell}}^2 + \noise} \right)\\
	&= \sum_{k = 1}^K  \log_2 \left( \frac{\tr{ \mF \mW \mW^\H \mF^\H \vh_k \vh_k^\H } + \noise}{\tr{\mF \mW_{\bar{k}} \mW_{\bar{k}}^\H \mF^\H \vh_k \vh_k^\H } + \noise}\right) \nbthis \label{eq_rate_11} \\
	&= \sum_{k = 1}^K  \log_2 \left(\tr{ \mF \mV \mF^\H \tilde{\mH}_k} + \noise \right)\\
	&\qquad -  \sum_{k = 1}^K \log_2 \left(\tr{\mF \mV_{\bar{k}} \mF^\H \tilde{\mH}_k } + \noise \right), \nbthis \label{eq_rate_2}
\end{align*}
where $\mW_{\bar{k}}$, $\mV$, $\mV_{\bar{k}}$, and $\tilde{\mH}_k$         are defined in Theorem \ref{theo_gradient_rate}. Based on \eqref{eq_rate_2}, we can compute $\nabla_{\mF} R$ as
\begin{align*}
	\hspace{-0.5cm} \nabla_{\mF} R &=  \sum_{k = 1}^K  \underbrace{\frac{\partial }{\partial \mF^*} \log_2 \left(\tr{ \mF \mV \mF^\H \tilde{\mH}_k} + \noise \right)}_{\triangleq \partial_{k1}} \\
	& \quad -  \sum_{k = 1}^K  \underbrace{\frac{\partial }{\partial \mF^*} \log_2 \left(\tr{\mF^\H \mV_{\bar{k}} \mF^\H \tilde{\mH}_k } + \noise \right)}_{\triangleq \partial_{k2}}. \nbthis \label{eq_derivation_1}
\end{align*}
Using the result that $\partial \tr{\mZ \mA_0 \mZ^\H \mA_1}/\partial \mZ^* = \mA_1 \mZ \mA_0$ in~\cite{hjorungnes2007complex}, we have
\begin{align*}
	\partial_{k1} &= \frac{\frac{\partial }{\partial \mF^*} \left(\tr{ \mF \mV \mF^\H \tilde{\mH}_k} + \noise \right)}{\ln2 (\tr{ \mF \mV \mF^\H \tilde{\mH}_k} + \noise )}\\
	&= \frac{\tilde{\mH}_k \mF \mV}{\ln2 (\tr{ \mF \mV \mF^\H \tilde{\mH}_k} + \noise )}, \nbthis \label{eq_derivation_2} 
\end{align*}
and similarly,
\begin{align*}
	\partial_{k2}= \frac{\tilde{\mH}_k \mF \mV_{\bar{k}}}{\ln2 (\tr{ \mF \mV_{\bar{k}} \mF^\H \tilde{\mH}_k} + \noise )}. \nbthis \label{eq_derivation_3} 
\end{align*}
Substituting \eqref{eq_derivation_2} and \eqref{eq_derivation_3} into \eqref{eq_derivation_1} yields \eqref{grad_rate_F} in Theorem \ref{theo_gradient_rate}.

To compute $\nabla_{\mW} R$, we write $R$ in \eqref{eq_rate_11} as
\begin{align*}
	R
	&= \sum_{k = 1}^K \log_2 \left( \frac{\tr{ \mW \mW^\H \mF^\H \vh_k \vh_k^\H \mF } + \noise}{\tr{ \mW_{\bar{k}} \mW_{\bar{k}}^\H \mF^\H \vh_k \vh_k^\H \mF} + \noise}\right) \\
	&= \sum_{k = 1}^K \log_2 \left(\tr{ \mW \mW^\H \bar{\mH}_k } + \noise \right)\\
	&\qquad - \sum_{k = 1}^K \log_2 \left(\tr{ \mW_{\bar{k}} \mW_{\bar{k}}^\H \bar{\mH}_k} + \noise \right), \nbthis \label{eq_rate_4}
\end{align*}
with $\bar{\mH}_k$ defined in \eqref{def_Hmbar}. Following similar derivations as in \eqref{eq_derivation_1}--\eqref{eq_derivation_3}, we obtain \eqref{grad_rate_W}, and the proof is completed.

\section{Proof of Theorem \ref{theo_gradient_tau}}

\label{app_proof_grad_tau}
The derivation of the gradients of $\tau$ with respect to $\mF$ and $\mW$, i.e., $\nabla_{\mF} \tau$ and $\nabla_{\mW} \tau$, is challenging. To tackle this, we first recall the following definitions:
\begin{align*}
	\nabla_{\mZ} f & = \frac{\partial f}{\partial \mZ^*} = \begin{bmatrix}
		\frac{\partial f}{\partial [\mZ]_{11}^*} & \ldots & \frac{\partial f}{\partial [\mZ]_{1C}^*}\\
		\vdots &   \vdots & \vdots \\
		\frac{\partial f}{\partial [\mZ]_{R1}^*} & \ldots & \frac{\partial f}{\partial [\mZ]_{R C}^*}
	\end{bmatrix}, \nbthis \label{eq_grad_tau_F_0}
\end{align*}
where $\mZ \in \setC^{R \times C}$. Thus, $\nabla_{\mF} \tau$ and $\nabla_{\mW} \tau$ can be obtained using $\partial \tau / \partial [\mF]_{nm}^*$ and $\partial \tau / \partial [\mW]_{mk}^*$, respectively, with $n = 1,\ldots,N$, $m = 1,\ldots,M$, and $k = 1,\ldots, K$. 

Let us denote $\mU \triangleq \mF \mW \mW^\H \mF^\H \in \setC^{N \times N}$ and rewrite $\tau$ as $\tau = \norm{ \mU - \bm{\Psi} }_{\mathcal{F}}^2$. Applying the following chain rule to $\tau$, $\partial \tau / \partial [\mF]_{nm}^*$ and $\partial \tau / \partial [\mW]_{mk}^*$ can be derived as follows
\begin{align*}
	\frac{\partial \tau}{\partial [\mF]_{nm}^*} &= \trt{\left( \frac{\partial \tau}{\partial \mU^\H} \right)^\T \frac{\partial \mU^\H}{\partial [\mF]_{nm}^*} } = \trt{\frac{\partial \tau}{\partial \mU^*} \frac{\partial \mU}{\partial [\mF]_{nm}^*} }, \nbthis \label{eq_grad_tau_F_2}\\
	\frac{\partial \tau}{\partial [\mW]_{mk}^*} &= \trt{\left( \frac{\partial \tau}{\partial \mU^\H} \right)^\T \frac{\partial \mU^\H}{\partial [\mW]_{mk}^*} } = \trt{\frac{\partial \tau}{\partial \mU^*} \frac{\partial \mU}{\partial [\mW]_{mk}^*} }, \nbthis \label{eq_grad_tau_W_2}
\end{align*}
where \eqref{eq_grad_tau_F_2} and \eqref{eq_grad_tau_W_2} follow from the fact that $\mU = \mU^\H$.

\subsection{Derivation of $\partial \tau/\partial \mU^*$}
Since both $\partial \tau/\partial \mF^*$ and $\partial \tau/\partial \mW^*$ depend on $\partial \tau/\partial \mU^*$ as seen in \eqref{eq_grad_tau_F_2} and \eqref{eq_grad_tau_W_2}, we first need to compute $\partial \tau/\partial \mU^*$. We rewrite
\begin{align*}
	\tau = \tr{\mU \mU^\H - \bm{\Psi} \mU^\H - \mU \bm{\Psi}^\H + \bm{\Psi} \bm{\Psi}^\H}, 
\end{align*}
and note that since $\partial \tr{\mU \mU^\H}/\partial \mU^* = 2 \mU$ and $\partial \tr{\mU \bm{\Psi}^\H}/\partial \mU^* = \partial \tr{\bm{\Psi} \mU^\H}/\partial \mU^* = \bm{\Psi}$ \cite{petersen2008matrix}, we have
\begin{align*}
	\frac{\partial \tau}{\partial \mU^*} = 2(\mU - \bm{\Psi}). \nbthis \label{eq_grad_tau_R}
\end{align*}

\subsection{Derivation of $\partial \tau/\partial \mF^*$}
We now compute $\partial \mU/\partial [\mF]_{nm}^*$ in \eqref{eq_grad_tau_F_2}. Let us write $[\mU]_{ij} = \tr{\bm{\delta}_i^\H \mF \mW \mW^\H \mF^\H \bm{\delta}_j } = \tr{\mF \mW \mW^\H \mF^\H \bm{\delta}_j \bm{\delta}_i^\H}$ where $\bm{\delta}_i$ and $\bm{\delta}_j$ are the $i$-th and $j$-th columns of identity matrix $\mI_N$, respectively. Then, using the result $ \partial \tr{\mZ \mA_0 \mZ^\H \mA_1} / \partial \mZ^* = \mA_1 \mZ \mA_0$ in \cite{hjorungnes2007complex}, we have
\begin{align*}
	\frac{\partial [\mU]_{ij}}{\partial \mF^*} = \bm{\delta}_j \bm{\delta}_i^\H \mF \mW \mW^\H.  \nbthis \label{eq_grad_tau_F_3}
\end{align*}
Furthermore, since $\partial [\mU]_{ij} / \partial [\mF]_{nm}^*$ is the $(n, m)$-th entry of $\partial [\mU]_{ij} / \partial \mF^*$, we can write
\begin{align*}
	\frac{\partial [\mU]_{ij}}{\partial [\mF]_{nm}^*} = \bm{\delta}_n^\H \bm{\delta}_j \bm{\delta}_i^\H \mF \mW \mW^\H \bm{\delta}_m = \bm{\delta}_i^\H \mF \mW \mW^\H \bm{\delta}_m \bm{\delta}_n^\H \bm{\delta}_j ,  \nbthis \label{eq_grad_tau_F_4}
\end{align*}
where $\bm{\delta}_n$ and $\bm{\delta}_m$ are the $n$-th and $m$-th columns of identity matrices $\mI_N$ and $\mI_{M}$, respectively. The second equality in \eqref{eq_grad_tau_F_4} holds because $\bm{\delta}_n^\H \bm{\delta}_j$ is a scalar. Thus, we have
\begin{align*}
	\frac{\partial \mU}{\partial [\mF]_{nm}^*} = \mF \mW \mW^\H \bm{\delta}_m \bm{\delta}_n^\H.  \nbthis \label{eq_grad_tau_F_5}
\end{align*}
Substituting \eqref{eq_grad_tau_R} and \eqref{eq_grad_tau_F_5} into \eqref{eq_grad_tau_F_2} yields
\begin{align*}
	\frac{\partial \tau}{\partial [\mF]_{nm}^*} &= 2 \tr{(\mU-\bm{\Psi}) \mF \mW \mW^\H \bm{\delta}_m \bm{\delta}_n^\H}\\
	&= 2 \bm{\delta}_n^\H(\mU-\bm{\Psi}) \mF \mW \mW^\H \bm{\delta}_m.  \nbthis \label{eq_grad_tau_F_6}
\end{align*}
Again, we utilize the fact that $\partial \tau/\partial [\mF]_{nm}^*$ is the $(n,m)$-th element of $\partial \tau/\partial [\mF]_{nm}^*$ to obtain
\begin{align*}
	\frac{\partial \tau}{\partial \mF^*} &= 2 (\mU-\bm{\Psi}) \mF \mW \mW^\H.  \nbthis \label{eq_grad_tau_F_7}
\end{align*}
Replacing $\mU$ by $\mF \mW \mW^\H \mF^\H$ in \eqref{eq_grad_tau_F_7} gives us the result \eqref{grad_tau_F}.

\subsection{Derivation of $\partial \tau/\partial \mW^*$}
The derivation of $\partial \tau/\partial \mW^*$ can be found in a similar manner. Specifically, we first write
\begin{align*}
	[\mU]_{ij} = \tr{\bm{\delta}_i^\H \mF \mW \mW^\H \mF^\H \bm{\delta}_j } = \tr{\mW \mW^\H \mF^\H \bm{\delta}_j \bm{\delta}_i^\H \mF }.
\end{align*}
Then, we apply the result $\partial \tr{\mZ \mA_0 \mZ^\H \mA_1} / \partial \mZ^* = \mA_1 \mZ \mA_0$ in \cite{hjorungnes2007complex} with $\mA_0 = \mI$ and $\mA_1 = \mF^\H \bm{\delta}_j \bm{\delta}_i^\H \mF$ to obtain
\begin{align*}
	\frac{\partial [\mU]_{ij}}{\partial \mW^*} &= \mF^\H \bm{\delta}_j \bm{\delta}_i^\H \mF \mW,  \nbthis \label{eq_grad_tau_W_3} \\
	\frac{\partial [\mU]_{ij}}{\partial [\mW]_{mk}^*} &= \bm{\delta}_m^\H \mF^\H \bm{\delta}_j \bm{\delta}_i^\H \mF \mW \bm{\delta}_k =  \bm{\delta}_i^\H \mF \mW \bm{\delta}_k \bm{\delta}_m^\H \mF^\H \bm{\delta}_j,  \nbthis \label{eq_grad_tau_W_4}
\end{align*}
which leads to
\begin{align*}
	\frac{\partial \mU}{\partial [\mW]_{mk}^*} = \mF \mW \bm{\delta}_k \bm{\delta}_m^\H \mF^\H.  \nbthis \label{eq_grad_tau_W_5}
\end{align*}
Substituting \eqref{eq_grad_tau_R} and \eqref{eq_grad_tau_W_5} into \eqref{eq_grad_tau_W_2} gives
\begin{align*}
	\frac{\partial \tau}{\partial [\mW]_{mk}^*} &= 2 \tr{(\mU-\bm{\Psi}) \mF \mW \bm{\delta}_k \bm{\delta}_m^\H \mF^\H}\\
	&= 2 \bm{\delta}_m^\H \mF^\H (\mU-\bm{\Psi}) \mF \mW \bm{\delta}_k,
\end{align*}
or equivalently,
\begin{align*}
	\frac{\partial \tau}{\partial \mW^*} = 2 \mF^\H (\mF \mW \mW^\H \mF^\H - \bm{\Psi}) \mF \mW,
\end{align*}
which is \eqref{grad_tau_W}, and the proof is completed.

\bibliographystyle{IEEEtran}
\bibliography{IEEEabrv,Bibliography}

\begin{thebibliography}{10}
\providecommand{\url}[1]{#1}
\csname url@samestyle\endcsname
\providecommand{\newblock}{\relax}
\providecommand{\bibinfo}[2]{#2}
\providecommand{\BIBentrySTDinterwordspacing}{\spaceskip=0pt\relax}
\providecommand{\BIBentryALTinterwordstretchfactor}{4}
\providecommand{\BIBentryALTinterwordspacing}{\spaceskip=\fontdimen2\font plus
\BIBentryALTinterwordstretchfactor\fontdimen3\font minus
  \fontdimen4\font\relax}
\providecommand{\BIBforeignlanguage}[2]{{%
\expandafter\ifx\csname l@#1\endcsname\relax
\typeout{** WARNING: IEEEtran.bst: No hyphenation pattern has been}%
\typeout{** loaded for the language `#1'. Using the pattern for}%
\typeout{** the default language instead.}%
\else
\language=\csname l@#1\endcsname
\fi
#2}}
\providecommand{\BIBdecl}{\relax}
\BIBdecl

\bibitem{giordani2020toward}
M.~Giordani, M.~Polese, M.~Mezzavilla, S.~Rangan, and M.~Zorzi, ``{Toward 6G
  networks: Use cases and technologies},'' \emph{{IEEE} Commun. Mag.}, vol.~58,
  no.~3, pp. 55--61, 2020.

\bibitem{samsung202065}
``{6G} - {T}he next hyper connected experience for all,'' \emph{Samsung 6G
  Vision}, 2020.

\bibitem{rappaport2019wireless}
T.~S. Rappaport, Y.~Xing, O.~Kanhere, S.~Ju, A.~Madanayake, S.~Mandal,
  A.~Alkhateeb, and G.~C. Trichopoulos, ``{Wireless communications and
  applications above 100 GHz: Opportunities and challenges for 6G and
  beyond},'' \emph{{IEEE} Access}, vol.~7, pp. 78\,729--78\,757, 2019.

\bibitem{nguyen2022beam}
{N. T. Nguyen \textit{et al.}}, ``Beam squint effects in {THz} communications
  with {UPA and ULA: C}omparison and hybrid beamforming design,'' in
  \emph{Proc. IEEE Global Commun. Conf. Workshop}, 2022.

\bibitem{mishra2019toward}
K.~V. Mishra, M.~B. Shankar, V.~Koivunen, B.~Ottersten, and S.~A. Vorobyov,
  ``{Toward millimeter-wave joint radar communications: A signal processing
  perspective},'' \emph{{IEEE} Signal Process. Mag.}, vol.~36, no.~5, pp.
  100--114, 2019.

\bibitem{molisch2017hybrid}
A.~F. Molisch, V.~V. Ratnam, S.~Han, Z.~Li, S.~L.~H. Nguyen, L.~Li, and
  K.~Haneda, ``Hybrid beamforming for massive {MIMO}: A survey,'' \emph{{IEEE}
  Commun. Mag.}, vol.~55, no.~9, pp. 134--141, 2017.

\bibitem{zhang2021overview}
J.~A. Zhang, F.~Liu, C.~Masouros, R.~W. Heath, Z.~Feng, L.~Zheng, and
  A.~Petropulu, ``An overview of signal processing techniques for joint
  communication and radar sensing,'' \emph{{IEEE} J. Sel. Topics Signal
  Process.}, vol.~15, no.~6, pp. 1295--1315, 2021.

\bibitem{ouyang2022performance}
C.~Ouyang, Y.~Liu, and H.~Yang, ``Performance of downlink and uplink integrated
  sensing and communications {(ISAC)} systems,'' \emph{{IEEE} Wireless Commun.
  Lett.}, vol.~11, no.~9, pp. 1850--1854, 2022.

\bibitem{liu2022integrated}
F.~Liu, Y.~Cui, C.~Masouros, J.~Xu, T.~X. Han, Y.~C. Eldar, and S.~Buzzi,
  ``Integrated sensing and communications: {T}owards dual-functional wireless
  networks for {6G} and beyond,'' \emph{{IEEE} J. Sel. Areas Commun.}, 2022.

\bibitem{zhang2018multibeam}
J.~A. Zhang, X.~Huang, Y.~J. Guo, J.~Yuan, and R.~W. Heath, ``Multibeam for
  joint communication and radar sensing using steerable analog antenna
  arrays,'' \emph{{IEEE} Trans. Veh. Technol.}, vol.~68, no.~1, pp. 671--685,
  2018.

\bibitem{ma2020joint}
D.~Ma, N.~Shlezinger, T.~Huang, Y.~Liu, and Y.~C. Eldar, ``Joint
  radar-communication strategies for autonomous vehicles: {C}ombining two key
  automotive technologies,'' \emph{{IEEE} Signal Process. Mag.}, vol.~37,
  no.~4, pp. 85--97, 2020.

\bibitem{huang2020majorcom}
T.~Huang, N.~Shlezinger, X.~Xu, Y.~Liu, and Y.~C. Eldar, ``{MAJoRCom}: A
  dual-function radar communication system using index modulation,''
  \emph{{IEEE} Trans. Signal Process.}, vol.~68, pp. 3423--3438, 2020.

\bibitem{ma2021spatial}
D.~Ma, N.~Shlezinger, T.~Huang, Y.~Shavit, M.~Namer, Y.~Liu, and Y.~C. Eldar,
  ``Spatial modulation for joint radar-communications systems: Design,
  analysis, and hardware prototype,'' \emph{{IEEE} Trans. Veh. Technol.},
  vol.~70, no.~3, pp. 2283--2298, 2021.

\bibitem{ma2021frac}
D.~Ma, N.~Shlezinger, T.~Huang, Y.~Liu, and Y.~C. Eldar, ``{FRaC}: {FMCW}-based
  joint radar-communications system via index modulation,'' \emph{{IEEE} J.
  Sel. Topics Signal Process.}, vol.~15, no.~6, pp. 1348--1364, 2021.

\bibitem{Hassanien2016Dual}
A.~{Hassanien}, M.~G. {Amin}, Y.~D. {Zhang}, and F.~{Ahmad}, ``Dual-function
  radar-communications: Information embedding using sidelobe control and
  waveform diversity,'' \emph{{IEEE} Trans. Signal Process.}, vol.~64, no.~8,
  pp. 2168--2181, 2016.

\bibitem{Kumari2018IEEE80211ad}
P.~{Kumari}, J.~{Choi}, N.~{González-Prelcic}, and R.~W. {Heath}, ``{IEEE}
  802.11ad-based radar: An approach to joint vehicular communication-radar
  system,'' \emph{{IEEE} Trans. Veh. Technol.}, vol.~67, no.~4, pp. 3012--3027,
  2018.

\bibitem{chepuri2022integrated}
S.~P. Chepuri, N.~Shlezinger, F.~Liu, G.~C. Alexandropoulos, S.~Buzzi, and
  Y.~C. Eldar, ``Integrated sensing and communications with reconfigurable
  intelligent surfaces,'' \emph{{IEEE} Signal Process. Mag.}, 2023, early
  access.

\bibitem{liu2018mu}
F.~Liu, C.~Masouros, A.~Li, H.~Sun, and L.~Hanzo, ``{MU-MIMO communications
  with MIMO radar: From co-existence to joint transmission},'' \emph{{IEEE}
  Trans. Wireless Commun.}, vol.~17, no.~4, pp. 2755--2770, 2018.

\bibitem{li2016optimum}
B.~Li, A.~P. Petropulu, and W.~Trappe, ``{Optimum co-design for spectrum
  sharing between matrix completion based MIMO radars and a MIMO communication
  system},'' \emph{{IEEE} Trans. Signal Process.}, vol.~64, no.~17, pp.
  4562--4575, 2016.

\bibitem{liu2022transmit}
X.~Liu, T.~Huang, Y.~Liu, and Y.~C. Eldar, ``Transmit beamforming with fixed
  covariance for integrated {MIMO} radar and multiuser communications,'' in
  \emph{Proc. IEEE Int. Conf. Acoust., Speech, Signal Processing}, 2022, pp.
  8732--8736.

\bibitem{pritzker2022transmit}
J.~Pritzker, J.~Ward, and Y.~C. Eldar, ``Transmit precoder design approaches
  for dual-function radar-communication systems,'' \emph{arXiv preprint
  arXiv:2203.09571}, 2022.

\bibitem{liu2020joint}
X.~Liu, T.~Huang, N.~Shlezinger, Y.~Liu, J.~Zhou, and Y.~C. Eldar, ``{Joint
  transmit beamforming for multiuser MIMO communications and MIMO radar},''
  \emph{{IEEE} Trans. Signal Process.}, vol.~68, pp. 3929--3944, 2020.

\bibitem{pritzker2021transmit}
J.~Pritzker, J.~Ward, and Y.~C. Eldar, ``Transmit precoding for dual-function
  radar-communication systems,'' in \emph{Proc. Annual Asilomar Conf. Signals,
  Syst., Comp.}, 2021.

\bibitem{liu2018toward}
F.~Liu, L.~Zhou, C.~Masouros, A.~Li, W.~Luo, and A.~Petropulu, ``Toward
  dual-functional radar-communication systems: Optimal waveform design,''
  \emph{{IEEE} Trans. Signal Process.}, vol.~66, no.~16, pp. 4264--4279, 2018.

\bibitem{tang2022mimo}
B.~Tang and P.~Stoica, ``{MIMO multifunction RF systems: Detection performance
  and waveform design},'' \emph{{IEEE} Trans. Signal Process.}, vol.~70, pp.
  4381--4394, 2022.

\bibitem{wu2023constant}
W.~Wu, B.~Tang, and X.~Wang, ``Constant-modulus waveform design for
  dual-function radar-communication systems in the presence of clutter,''
  \emph{{IEEE} Trans. Aerosp. Electron. Syst.}, 2023.

\bibitem{liu2021dual}
R.~Liu, M.~Li, Q.~Liu, and A.~L. Swindlehurst, ``{Dual-functional
  radar-communication waveform design: A symbol-level precoding approach},''
  \emph{{IEEE} J. Sel. Topics Signal Process.}, vol.~15, no.~6, pp. 1316--1331,
  2021.

\bibitem{liu2022joint}
------, ``Joint waveform and filter designs for {STAP-SLP}-based {MIMO-DFRC}
  systems,'' \emph{{IEEE} J. Sel. Areas Commun.}, vol.~40, no.~6, pp.
  1918--1931, 2022.

\bibitem{wu2022joint}
K.~Wu, J.~A. Zhang, Z.~Ni, X.~Huang, Y.~J. Guo, and S.~Chen, ``Joint
  communications and sensing employing optimized {MIMO-OFDM} signals,''
  \emph{arXiv preprint arXiv:2208.09791}, 2022.

\bibitem{johnston2022mimo}
J.~Johnston, L.~Venturino, E.~Grossi, M.~Lops, and X.~Wang, ``{MIMO OFDM
  dual-function radar-communication under error rate and beampattern
  constraints},'' \emph{{IEEE} J. Sel. Areas Commun.}, vol.~40, no.~6, pp.
  1951--1964, 2022.

\bibitem{temiz2021dual}
M.~Temiz, E.~Alsusa, and M.~W. Baidas, ``{A dual-function massive MIMO uplink
  OFDM communication and radar architecture},'' \emph{{IEEE} Trans. on Cogn.
  Commun. Netw.}, vol.~8, no.~2, pp. 750--762, 2021.

\bibitem{buzzi2019using}
S.~Buzzi, C.~D’Andrea, and M.~Lops, ``{Using massive MIMO arrays for joint
  communication and sensing},'' in \emph{Proc. Annual Asilomar Conf. Signals,
  Syst., Comp.}, 2019, pp. 5--9.

\bibitem{keskin2021mimo}
M.~F. Keskin, H.~Wymeersch, and V.~Koivunen, ``{MIMO-OFDM joint
  radar-communications: Is ICI friend or foe?}'' \emph{{IEEE} J. Sel. Topics
  Signal Process.}, vol.~15, no.~6, pp. 1393--1408, 2021.

\bibitem{zirtiloglu2022power}
T.~Zirtiloglu, N.~Shlezinger, Y.~C. Eldar, and R.~T. Yazicigil,
  ``Power-efficient hybrid {MIMO} reciever with task-specific beamforming using
  low-resolution {ADCs},'' in \emph{Proc. IEEE Int. Conf. Acoust., Speech,
  Signal Processing}, 2022, pp. 5338--5342.

\bibitem{shlezinger2021dynamic}
N.~Shlezinger, G.~C. Alexandropoulos, M.~F. Imani, Y.~C. Eldar, and D.~R.
  Smith, ``Dynamic metasurface antennas for {6G} extreme massive {MIMO}
  communications,'' \emph{{IEEE} Wireless Commun.}, vol.~28, no.~2, pp.
  106--113, 2021.

\bibitem{zhang2022holographic}
H.~Zhang, H.~Zhang, B.~Di, M.~Di~Renzo, Z.~Han, H.~V. Poor, and L.~Song,
  ``Holographic integrated sensing and communication,'' \emph{{IEEE} J. Sel.
  Areas Commun.}, vol.~40, no.~7, pp. 2114--2130, 2022.

\bibitem{gong2019rf}
T.~Gong, N.~Shlezinger, S.~S. Ioushua, M.~Namer, Z.~Yang, and Y.~C. Eldar,
  ``{RF} chain reduction for {MIMO} systems: A hardware prototype,''
  \emph{{IEEE} Syst. J.}, vol.~14, no.~4, pp. 5296--5307, 2020.

\bibitem{qi2022hybrid}
C.~Qi, W.~Ci, J.~Zhang, and X.~You, ``Hybrid beamforming for millimeter wave
  {MIMO} integrated sensing and communications,'' \emph{{IEEE} Commun. Lett.},
  vol.~26, no.~5, pp. 1136--1140, 2022.

\bibitem{wang2022partially}
X.~Wang, Z.~Fei, J.~A. Zhang, and J.~Xu, ``Partially-connected hybrid
  beamforming design for integrated sensing and communication systems,''
  \emph{{IEEE} Trans. Commun.}, vol.~70, no.~10, pp. 6648--6660, 2022.

\bibitem{liyanaarachchi2021joint}
S.~D. Liyanaarachchi, C.~B. Barneto, T.~Riihonen, M.~Heino, and M.~Valkama,
  ``Joint multi-user communication and {MIMO} radar through full-duplex hybrid
  beamforming,'' in \emph{IEEE Int. Online Symposium Joint Commun. \& Sensing
  (JC\&S)}, 2021.

\bibitem{barneto2021beamformer}
C.~B. Barneto, T.~Riihonen, S.~D. Liyanaarachchi, M.~Heino,
  N.~González-Prelcic, and M.~Valkama, ``Beamformer design and optimization
  for joint communication and full-duplex sensing at mm-{W}aves,'' \emph{IEEE
  Trans. Commun.}, vol.~70, no.~12, pp. 8298--8312, Dec. 2022.

\bibitem{cheng2022QoS}
Z.~Cheng and B.~Liao, ``{Q}o{S}-aware hybrid beamforming and {DOA} estimation
  in multi-carrier dual-function radar-communication systems,'' \emph{IEEE J.
  Select. Areas in Commun.}, vol.~40, no.~6, pp. 1890--1905, Sept. 2022.

\bibitem{cheng2021hybrid_narrow}
Z.~Cheng, Z.~He, and B.~Liao, ``Hybrid beamforming for multi-carrier
  dual-function radar-communication system,'' \emph{{IEEE} Trans. on Cogn.
  Commun. Netw.}, vol.~7, no.~3, pp. 1002--1015, 2021.

\bibitem{wang2022HBD_OFDM}
B.~Wang, Z.~Cheng, L.~Wu, and Z.~He, ``Hybrid beamforming design for {OFDM}
  dual-function radar-communication system with double-phase-shifter
  structure,'' in \emph{Proc. European Signal Process. Conf.}, Belgrade,
  Serbia, Aug. 2022, pp. 1067--1071.

\bibitem{islam2022integrated}
M.~A. Islam, G.~C. Alexandropoulos, and B.~Smida, ``Integrated sensing and
  communication with millimeter wave full duplex hybrid beamforming,'' in
  \emph{Proc. IEEE Int. Conf. Commun.}, 2022, pp. 4673--4678.

\bibitem{cheng2021hybrid}
Z.~Cheng, Z.~He, and B.~Liao, ``Hybrid beamforming design for {OFDM}
  dual-function radar-communication system,'' \emph{{IEEE} J. Sel. Topics
  Signal Process.}, vol.~15, no.~6, pp. 1455--1467, 2021.

\bibitem{Elbir2021HB_THz}
A.~M. Elbir, K.~V. Mishra, and S.~Chatzinotas, ``Hybrid beamforming for
  {T}erahertz joint ultra-massive {MIMO} radar-communications,'' in \emph{Proc.
  Int. Symp. on Wireless Commun. Systems}, Berlin, Germany, Sept. 2021.

\bibitem{Kaushik2021Hardware}
A.~Kaushik, C.~Masouros, and F.~Liu, ``Hardware efficient joint
  radar-communications with hybrid precoding and {RF} chain optimization,'' in
  \emph{Proc. IEEE Int. Conf. Commun.}, Montreal, QC, Canada, June 2021.

\bibitem{Kaushik2022Green}
A.~Kaushik, E.~Vlachos, C.~Masouros, C.~Tsinos, and J.~Thompson, ``Green joint
  radar-communications: {RF} selection with low resolution {DAC}s and hybrid
  precoding,'' in \emph{Proc. IEEE Int. Conf. Commun.}, Seoul, Republic of
  Korea, May 2022, pp. 3160--3165.

\bibitem{liu2019hybrid}
F.~Liu and C.~Masouros, ``{Hybrid beamforming with sub-arrayed MIMO radar:
  Enabling joint sensing and communication at mmWave band},'' in \emph{Proc.
  IEEE Int. Conf. on Acoustics, Speech and Signal Process.}, Brighton, UK, May
  2019, pp. 7770--7774.

\bibitem{shlezinger2023AI}
N.~Shlezinger, M.~Ma, O.~Lavi, N.~T. Nguyen, Y.~C. Eldar, and M.~Juntti,
  ``{AI}-empowered hybrid {MIMO} beamforming,'' \emph{arXiv preprint
  arXiv:2303.01723}, 2023.

\bibitem{Mateos2022EndtoEnd}
J.~M. Mateos-Ramos, J.~Song, Y.~Wu, C.~Häger, M.~F. Keskin, V.~Yajnanarayana,
  and H.~Wymeersch, ``End-to-end learning for integrated sensing and
  communication,'' in \emph{Proc. IEEE Int. Conf. Commun.}, Seoul, Republic of
  Korea, May 2022, pp. 1942--1947.

\bibitem{muth2023Autoencoder}
C.~Muth and L.~Schmalen, ``Autoencoder-based joint communication and sensing of
  multiple targets,'' in \emph{Proc. Int. ITG Workshop on Smart Antennas and
  Conf. on Systems, Commun., and Coding}, Feb. 2023.

\bibitem{xu2022deep}
L.~Xu, R.~Zheng, and S.~Sun, ``A deep reinforcement learning approach for
  integrated automotive radar sensing and communication,'' in \emph{Proc. IEEE
  Sensor Array and Multichannel Sign. Proc. Workshop}, 2022, pp. 316--320.

\bibitem{elbir2021terahertz}
A.~M. Elbir, K.~V. Mishra, and S.~Chatzinotas, ``Terahertz-band joint
  ultra-massive {MIMO} radar-communications: {M}odel-based and model-free
  hybrid beamforming,'' \emph{{IEEE} J. Sel. Topics Signal Process.}, vol.~15,
  no.~6, pp. 1468--1483, 2021.

\bibitem{shlezinger2020model}
N.~Shlezinger, J.~Whang, Y.~C. Eldar, and A.~G. Dimakis, ``Model-based deep
  learning,'' \emph{Proc. {IEEE}}, 2023, early access.

\bibitem{monga2021algorithm}
V.~Monga, Y.~Li, and Y.~C. Eldar, ``Algorithm unrolling: {I}nterpretable,
  efficient deep learning for signal and image processing,'' \emph{{IEEE}
  Signal Process. Mag.}, vol.~38, no.~2, pp. 18--44, 2021.

\bibitem{shlezinger2022model}
N.~Shlezinger, J.~Whang, Y.~C. Eldar, and A.~G. Dimakis, ``Model-based deep
  learning,'' \emph{Proc. {IEEE}}, vol. 111, no.~5, pp. 465--499, 2023.

\bibitem{lavi2023learn}
O.~Lavi and N.~Shlezinger, ``Learn to rapidly and robustly optimize hybrid
  precoding,'' \emph{arXiv preprint arXiv:2301.00369}, 2023.

\bibitem{nguyen2023deep}
N.~T. Nguyen, M.~Ma, O.~Lavi, N.~Shlezinger, Y.~C. Eldar, A.~L. Swindlehurst,
  and M.~Juntti, ``Deep unfolding hybrid beamforming designs for {THz} massive
  {MIMO} systems,'' \emph{arXiv preprint arXiv:2302.12041}, 2023.

\bibitem{shlezinger2022discriminative}
N.~Shlezinger and T.~Routtenberg, ``Discriminative and generative learning for
  linear estimation of random signals [lecture notes],'' \emph{{IEEE} Signal
  Process. Mag.}, 2023, early access.

\bibitem{yu2016alternating}
X.~Yu, J.-C. Shen, J.~Zhang, and K.~B. Letaief, ``Alternating minimization
  algorithms for hybrid precoding in millimeter wave {MIMO} systems,''
  \emph{{IEEE} J. Sel. Topics Signal Process.}, vol.~10, no.~3, pp. 485--500,
  2016.

\bibitem{sohrabi2016hybrid}
F.~Sohrabi and W.~Yu, ``Hybrid digital and analog beamforming design for
  large-scale antenna arrays,'' \emph{{IEEE} J. Sel. Topics Signal Process.},
  vol.~10, no.~3, 2016.

\bibitem{nguyen2019unequally}
N.~T. Nguyen and K.~Lee, ``Unequally sub-connected architecture for hybrid
  beamforming in massive {MIMO} systems,'' \emph{{IEEE} Trans. Wireless
  Commun.}, vol.~19, no.~2, pp. 1127--1140, 2019.

\bibitem{boyd2004convex}
S.~Boyd and L.~Vandenberghe, \emph{Convex optimization}.\hskip 1em plus 0.5em
  minus 0.4em\relax Cambridge, 2004.

\bibitem{el2014spatially}
{O. El Ayach \textit{et al.}}, ``Spatially sparse precoding in millimeter wave
  {MIMO} systems,'' \emph{{IEEE} Trans. Wireless Commun.}, vol.~13, no.~3, pp.
  1499--1513, 2014.

\bibitem{agiv2022learn}
O.~Agiv and N.~Shlezinger, ``Learn to rapidly optimize hybrid precoding,'' in
  \emph{Proc. IEEE Works. on Sign. Proc. Adv. in Wirel. Comms.}, 2022.

\bibitem{liang2014low}
L.~Liang, W.~Xu, and X.~Dong, ``Low-complexity hybrid precoding in massive
  multiuser {MIMO} systems,'' \emph{{IEEE} Commun. Lett.}, vol.~3, no.~6, pp.
  653--656, 2014.

\bibitem{tran2012fast}
L.-N. Tran, M.~F. Hanif, A.~Tolli, and M.~Juntti, ``Fast converging algorithm
  for weighted sum rate maximization in multicell {MISO} downlink,''
  \emph{{IEEE} Signal Process. Lett.}, vol.~19, no.~12, pp. 872--875, 2012.

\bibitem{zhang2018performance}
C.~Zhang, Y.~Jing, Y.~Huang, and L.~Yang, ``Performance analysis for massive
  {MIMO} downlink with low complexity approximate zero-forcing precoding,''
  vol.~66, no.~9, pp. 3848--3864, 2018.

\bibitem{hjorungnes2007complex}
A.~Hjorungnes and D.~Gesbert, ``Complex-valued matrix differentiation:
  {T}echniques and key results,'' \emph{{IEEE} Trans. Signal Process.},
  vol.~55, no.~6, pp. 2740--2746, 2007.

\bibitem{petersen2008matrix}
K.~B. Petersen, M.~S. Pedersen \emph{et~al.}, ``The matrix cookbook,''
  \emph{Technical University of Denmark}, vol.~7, no.~15, p. 510, 2008.

\end{thebibliography}
\end{document}